\documentclass[aps,showpacs,preprintnumbers,amsmath,amssymb]{revtex4}
\usepackage{dcolumn}
\usepackage[dvips]{epsfig}
\usepackage{graphicx}
\usepackage{amssymb}
\usepackage{color}
\usepackage{enumerate}
\usepackage{subfigure}
\usepackage{multirow}
\usepackage{diagbox}

\usepackage{graphics,epsfig}
\usepackage{graphicx}

\begin{document}
\baselineskip=0.8 cm
%\preprint{example}
\title{\bf Shadow of a rotating squashed Kaluza-Klein black hole}
\author{Fen Long$^{1}$, Jieci Wang$^{1}$\footnote{jcwang@hunnu.edu.cn}, Songbai Chen$^{1, 2}$\footnote{csb3752@hunnu.edu.cn},  Jiliang Jing$^{1, 2}$\footnote{jljing@hunnu.edu.cn}}
%\email{csb3752@163.com}

\affiliation{$^{\textit{1}}$Institute of Physics and Department of Physics,  Key Laboratory of Low Dimensional Quantum Structures and Quantum Control of Ministry of Education, Synergetic Innovation Center for Quantum Effects and Applications,
Hunan Normal University,  Changsha,  Hunan 410081,  People's Republic
of China\\
$ ^2$Center for Gravitation and Cosmology, College of Physical Science and Technology,
Yangzhou University, Yangzhou 225009, China}

\begin{abstract}
\baselineskip=0.5 cm
\begin{center}
{\bf Abstract}
\end{center}

 We study the shadow of a rotating squashed Kaluza-Klein (KK) black hole and the shadow is found to possess distinct properties from those of usual rotating black holes. It is shown that  the shadow for a rotating squashed KK black hole is heavily influenced by the specific angular momentum of photon from the fifth dimension. Especially, as the parameters lie in a certain special range, there is no any shadow for a black hole, which does not emerge for the usual black holes. In the case where the black hole shadow exists, the shadow shape is a perfect black disk  and its radius decreases with the rotation parameter of the  black hole. Moreover,  the change of the shadow radius with extra dimension parameter also depends on the rotation parameter of black hole. Finally, with the latest observation data, we estimate  the angular radius of the shadow for the supermassive black hole Sgr $A^{*}$ at the centre of the Milky Way galaxy and the supermassive black hole in $M87$.

\end{abstract}

\maketitle
\newpage

\section{Introduction}

Recently, the Event Horizon Telescope team showcased the first image of the supermassive black hole in the center of the giant elliptical galaxy M87 \cite{e1,e001,e002,e003,e2,e3}. This event is terribly exciting  because it confirms once again that there exists exactly black hole in our Universe. Moreover, the information carried by the image is of great benefit to the understanding of the black hole shadow and the matter accretion process. Black hole shadow is a two-dimensional dark region in the observer's sky, where light rays from  the source fall into an event horizon. It is well known that the shape and size of shadow  depend on the black hole parameters \cite{p7,p8}, which implies that the shadow could be regarded as a potential tool to identify black holes. Thus, the shadows of black holes with different parameters have been studied recently in various theories of gravity \cite{p7,p8,ps9,p9,sw,swo,astro,chaotic,binary, sha18,my,sMN, w4,w5,w6,w7,w7s,w8,w9,w10,w11,w12,w13,w14,w15,w16,w17,w18,w19,w20,w21,w22,w23,w24,w25,Konoplya:2019sns, Giddings:2016btb,zt1}.
For example, the shadow  is a perfect rounded silhouette for a Schwarzschild black hole, but it becomes a ``D"-shaped silhouette for a fast rotating black hole due to  dragging effect \cite{p7,p8}. Moreover, the cusp silhouettes of shadows with small eye lashes are found in the spacetime of Kerr black hole with Proca hair \cite{ps9} and the  Konoplya-Zhidenko rotating non-Kerr black hole \cite{p9}. The self-similar fractal structures appear in the black hole shadows when the
equations of photon motion are not variable separable in the background spacetimes due to the chaotic motion of photon \cite{sw,swo,astro,chaotic,binary, sha18,my,sMN}. Recently, the shadows have been investigated for high-dimensional black holes including the Schwarzschild-Tangherlini black hole \cite{p10}, the five-dimensional rotating Myers-Perry black hole \cite{p13} and the five-dimensional rotating  Einstein-Maxwell-Chern-Simons black hole \cite{p14},  and the rotating black hole in Randall-Sundrum
theories \cite{p151}. These studies show that the extra dimension imprints in the black hole shadows.

Here, we focus on the  rotating squashed KK black hole, which is   a kind of  interesting Kaluza-Klein type metrics with the special topology and asymptotical structure \cite{j1}. This family of black holes
have squashed $S^3$ horizons. In the vicinity of horizon,  the black hole
has a structure like a five-dimensional black hole, but it behaves as the four-dimensional black holes with a constant twisted $S^1$ fiber in the far region. In these squashed KK black holes, the size of compactified extra dimension can be adjusted by the parameter $r_{\infty}$.  Recent investigations showed that the information of the size of the extra dimension for a KK black hole with squashed horizons imprints in its spectrum of Hawking radiation  \cite{j02,j2}, quasinormal frequencies \cite{j201,j202}, precession of a gyroscope in a circular orbit \cite{p19} and strong gravitational lensing \cite{p15}, which could open a possible window to observe extra dimensions in the future.  In the present work we study the shadow of a rotating squashed KK black hole and probe the features of the black hole shadow caused by the fifth dimension.

The paper is organized as follows. In Sec. $II$, we introduce briefly the rotating squashed Kaluza-Klein black hole spacetime and the corresponding null geodesics. In
Sec. $III$, we study the features in the shadow of a rotating squashed Kaluza-Klein black hole and find that the shadow is heavily influenced by the specific angular momentum of photon from the fifth dimension.
Finally, we present a summary and some discussions in the last section.

\section{The rotating squashed Kaluza-Klein black hole spacetime and null geodesics }

 The neutral rotating squashed KK black hole is a vacuum axisymmetric solution of Einstein field equation, which can be obtained by applying the squashing transformation technique to a five-dimensional Kerr black hole with two equal angular momenta \cite{j1}. The metric of this rotating squashed KK black hole can be expressed as
\begin{eqnarray}
d s^{2}=-d \tilde{t}^{\ 2}+\frac{\Sigma_{0}}{\Delta_{0}} k(r)^{2} dr^{2}
+\frac{r^{2}+a^{2}}{4}\left[k(r)(\sigma_{1}^{2}
+\sigma_{2}^{2})+\sigma_{3}^{2}\right]+\frac{M}{r^{2}+a^{2}}(d \tilde{t}-\frac{a}{2} \sigma_{3})^{2},\label{metric0}
\end{eqnarray}
with
\begin{eqnarray}
\Sigma_{0} &=&r^{2}(r^{2}+a^{2}), \\
\Delta_{0} &=&(r^{2}-r_{+}^{2})(r^{2}-r_{-}^{2}), \\
k(r)&=&\frac{(r_{\infty}^{2}-r_{+}^{2})(r_{\infty}^{2}-r_{-}^{2})}
{(r_{\infty}^{2}-r^{2})^{2}},\label{metric02}
\end{eqnarray}
and
\begin{eqnarray}
\sigma_{1} &=&-\sin \tilde{\psi} d \theta+\cos \tilde{\psi} \sin \theta d \phi,
 \\ \sigma_{2} &=&\cos \tilde{\psi} d \theta+\sin \tilde{\psi} \sin \theta d \phi, \\
\sigma_{3} &=&d \tilde{\psi}+\cos \theta d \phi, \label{metric01}
\end{eqnarray}
where the angular coordinates satisfy $0<\theta<\pi, 0<\phi<2 \pi $ and $ 0<\tilde{\psi}<4 \pi$.  The quantity $r_\infty$ corresponds to the spatial infinity and the polar coordinate $r$ runs in the range $\ 0<r<r_\infty$. The parameters $M$ and $a$ are associated with the mass and angular momenta of  the black hole, respectively. The outer and inner horizons of the black hole are located at $r_{+}$ and $r_{-}$,  which are functions of $M$ and $a$, i.e., $r^2_{\pm}=M-2a^2\pm\sqrt{M^2-4a^2M}$. The squashed parameter $k(r_{+})$ deforms the shape of black hole horizon.

Introducing a radial coordinate  \cite{j1}
\begin{eqnarray}
\rho=\tilde{\rho}_{0} \frac{r^{2}}{r_{\infty}^{2}-r^{2}},\label{new coordinate1}
\end{eqnarray}
with
\begin{eqnarray}
\tilde{\rho}_{0}^{2}=\frac{(r_{\infty}^{2}+a^{2})[(r_{\infty}^{2}+a^{2})^{2}-M r_{\infty}^{2}]}{4 r_{\infty}^{4}}.\label{new coordinate11}
\end{eqnarray}
One can find that the metric (\ref{metric0}) can be rewritten as
\begin{eqnarray}
d s^{2}=-d \tilde{t}^{\ 2}+U d \rho^{2}+R^{2}(\sigma_{1}^{2}+\sigma_{2}^{2})+W^{2} \sigma_{3}^{ 2}+V(d \tilde{t}-\frac{a}{2} \sigma_{3})^{2},\label{metric2}
\end{eqnarray}
with
\begin{eqnarray}
\begin{aligned}
K^{2}=\frac{\rho+\tilde{\rho}_{0}}{\rho+\frac{a^{2}}{r_{\infty}^{2}+a^{2}} \tilde{\rho}_{0}},\;\;\;\;\;\;\;
& V=\frac{M}{r_{\infty}^{2}+a^{2}} K^{2}, \;\;\;\;\;\;\;\quad W^{2}=\frac{r_{\infty}^{2}+a^{2}}{4 K^{2}}, \\ R^{2}=\frac{\left(\rho+\tilde{\rho}_{0}\right)^{2}}{K^{2}},\;\;\;\;\;\;\;\;\;\;\;\;\;\;
& U=\left(\frac{r_{\infty}^{2}}{r_{\infty}^{2}+a^{2}}\right)^{2} \frac{\tilde{\rho}_{0}^{2}}{W^{2}-\frac{r_{\infty}^{2}}{4} \frac{\rho}{\rho+\tilde{\rho}_{0}} V}.
\end{aligned}\label{metric21}
\end{eqnarray}
As the rotation parameter $a$ vanishes, the metric (\ref{metric2}) reduces to that of a five-dimensional Schwarzschild black hole with squashed horizon. When $r_{\infty}\rightarrow \infty $ ($k(r)\rightarrow 1$),  the squashing effect disappears and then the metric of the usual five-dimensional Kerr black hole with two equal angular momenta is recovered in this limit.
Adopting to the coordinates transformation \cite{j1}
\begin{eqnarray}
\tilde{t}=h\ t, \quad \quad \quad \tilde{\psi}=\psi-j\ t, \label{new coordinate2}
\end{eqnarray}
with
\begin{eqnarray}
h=\sqrt{\frac{\left(r_{\infty}^{2}+a^{2}\right)^{2}-M r_{\infty}^{2}}{\left(r_{\infty}^{2}+a^{2}\right)^{2}+M a^{2}}}, \quad \quad
j=\frac{2 M a}{\left(r_{\infty}^{2}+a^{2}\right)^{2}+M a^{2}}, \label{new coordinate21}
\end{eqnarray}
it is found that  the cross-term between $d \tilde{t}$ and $\sigma_{3}$ in the asymptotic form (i.e., $\rho \rightarrow \infty $) of the metric (\ref{metric2}) vanishes, which means that
 the asymptotic topology of the spacetime (\ref{metric2})   has the same form as the Schwarzschild squashed KK black hole spacetime. From the Komar mass $M_k$ for the black hole (\ref{metric2}) \cite{p15}
\begin{eqnarray}
\begin{aligned}
M_{k} &=\frac{M \pi}{2 G_{5}} \frac{(r_{\infty}^{2}+a^{2})^{2}-
M a^{2}}{\sqrt{(r_{\infty}^{2}+a^{2})^{2}+M a^{2}} \sqrt{(r_{\infty}^{2}+a^{2})^{2}-M r_{\infty}^{2}}} \\
&=\frac{M}{4 G_{4}} \frac{(r_{\infty}^{2}+a^{2})^{2}-M a^{2}}
{(r_{\infty}^{2}+a^{2})^{2}+M a^{2}} \frac{\sqrt{r_{\infty}^{2}+a^{2}}}{\sqrt{(r_{\infty}^{2}+a^{2})^{2}-M r_{\infty}^{2}}}.
\end{aligned}\label{new coordinate22}
\end{eqnarray}
we obtain the relationship between the five-dimensional and four-dimensional gravitational constants
\begin{eqnarray}
G_{5}=2 \pi r_{\infty}^{\prime} G_{4}, \label{new coordinate23}
\end{eqnarray}
with
\begin{eqnarray}
r_{\infty}^{\prime}=\sqrt{\frac{\left(r_{\infty}^{2}+a^{2}\right)^{2}+M a^{2}}{r_{\infty}^{2}+a^{2}}}. \label{new coordinate24}
\end{eqnarray}
Obviously, the expression of $r_{\infty}^{\prime}$ is more complicated than that of $r_{\infty}$ . However,  the geometric interpretation  for $r_{\infty}^{\prime}$ is clearer  than that of  $r_{\infty}$ in the rotating squashed KK black hole spacetime (\ref{metric2}) \cite{j1},  which implies that the parameter $r_{\infty}^{\prime}$  is better than $r_{\infty}$ for the compactified dimension. As the rotation parameter $a$ disappears, we find that $r_{\infty}^{\prime}$ reduces to $r_{\infty}$ and then the relationship (\ref{new coordinate23}) tends to the usual form in the Schwarzschild squashed KK black hole spacetime, i.e., $G_{5}=2\pi r_{\infty} G_{4}$. As in Ref. \cite{p19}, one can introduce a quantity $\rho_{M}$, which is related to Komar mass $M_{k}$ by
\begin{eqnarray}
\rho_{M}\equiv2 G_{4} M_{k}=\frac{M}{2} \frac{(r_{\infty}^{2}+a^{2})^{2}-M a^{2}}{(r_{\infty}^{2}+a^{2})^{2}
+M a^{2}} \frac{\sqrt{r_{\infty}^{2}+a^{2}}}{\sqrt{(r_{\infty}^{2}+a^{2})^{2}-M r_{\infty}^{2}}}.
\label{new coordinate25}
\end{eqnarray}
For a sake of simplifying the calculation, we make use of  the following transformations
\begin{eqnarray}
r^{\prime 2}=\frac{(r_{\infty}^{2}+a^{2})^{2}+M a^{2}}{r_{\infty}^{4}+r^{2} a^{2}} r^{2},\quad\quad \quad\rho_{0}^{2}=\frac{r_{\infty}^{\prime 2}-M}{4},\quad\quad \quad b=\sqrt{\frac{M}{r_{\infty}^{2}+a^{2}}} \ a,
\label{new coordinate26}
\end{eqnarray}
and then find that the radial coordinate (\ref{new coordinate1}) and the quantity $\rho_M$ can be rewritten as
\begin{eqnarray}
\rho&=&\rho_{0} \frac{r^{\prime 2}}{r_{\infty}^{\prime 2}-r^{\prime 2}},\quad\quad \quad\nonumber\\ \rho_{M}&=&\frac{\rho_{0} M}{r_{\infty}^{\prime 2}-M} \frac{(r_{\infty}^{2}+a^{2})^{2}-M a^{2}}
{(r_{\infty}^{2}+a^{2})^{2}+M a^{2}}=\frac{M \rho_{0}}{r_{\infty}^{\prime 2}-M}
(1-\frac{2 b^{2}}{r_{\infty}^{\prime 2}}).
\label{new coordinate29}
\end{eqnarray}
With these transformations, one can find that the metric (\ref{metric2}) can be given in a new form  \cite{p15}
\begin{eqnarray}
d s^{2}=-A(\rho) d t^{2}+B(\rho) d \rho^{2}+C(\rho)(d \theta^{2}+\sin ^{2} \theta d \phi^{2})+D(\rho)(d \psi+\cos \theta d \phi)^{2}-2 H(\rho) d t(d \psi+\cos \theta d \phi), \label{metric1}
\end{eqnarray}
where
\begin{eqnarray}
\begin{aligned} A(\rho) &=\frac{4-\mathcal{V}(a j-2)^{2}-4 j^{2} \mathcal{W}^{2}}{4 h^{2}}, \quad\quad \quad B(\rho)=\mathcal{U}(\rho), \quad\quad\quad C(\rho)=\mathcal{R}^{2}(\rho), \\
D(\rho) &=\frac{a^{2} \mathcal{V}+4 \mathcal{W}^{2}}{4}, \quad\quad\quad\quad H(\rho)=\frac{2 a \mathcal{V}-(a^{2} \mathcal{V}+4 \mathcal{W}^{2}) j}{4 h}, \end{aligned} \label{metric11}
\end{eqnarray}
with
\begin{eqnarray}
\begin{array}{c}{\mathcal{K}^{2}=\frac{\rho+\frac{r_{\infty}^{2}+a^{2}}{r_{\infty}^{2}} \rho_{0}}{\rho+\frac{a^{2}}{r_{\infty}^{2}} \rho_{0}}, \quad\quad\quad \mathcal{V}=\frac{M}{r_{\infty}^{2}+a^{2}} \mathcal{K}^{2}, \quad\quad\quad \mathcal{W}^{2}=\frac{r_{\infty}^{2}+a^{2}}{4 \mathcal{K}^{2}}}, \\ {\mathcal{R}^{2}(\rho)=(\rho+\frac{a^{2}}{r_{\infty}^{2}} \rho_{0})(\rho+\frac{r_{\infty}^{2}+a^{2}}{r_{\infty}^{2}} \rho_{0}), \quad\quad\quad\quad \mathcal{U}(\rho)=\frac{\rho_{0}^{2}}{\mathcal{W}^{2}-\frac{r_{\infty}^{2}}{4} \frac{\rho}{\mathcal{K} \mathcal{R}} \mathcal{V}}}.\end{array} \label{metric12}
\end{eqnarray}
There are only three independent parameters among the parameters $\rho_{0}$, $M$, $a$, $r_{\infty}$, $r_{\infty}^{\prime}$, $\rho_{M}$ and $b$. Here, we select the parameters  $\rho_0$, $\rho_M $ and $\ b$ as the independent parameters and the others are related to  them by
\begin{eqnarray}
\begin{aligned}
r_{\infty}^{2}\ = \ &2 \rho_{0}(\rho_{M}+\rho_{0})-b^{2}+\sqrt{b^{4}-4 b^{2} \rho_{0}(\rho_{0}-\rho_{M})+4 \rho_{0}^{2}(\rho_{M}+\rho_{0})^{2}},\\
&-\frac{b^{2}(4\rho_{0}^{2}-b^{2})}{2\rho_{0}(\rho_{M}-\rho_{0})+b^{2}+\sqrt{b^{4}-4 b^{2} \rho_{0}(\rho_{0}-\rho_{M})+4 \rho_{0}^{2}(\rho_{M}+\rho_{0})^{2}}},\\
r_{\infty}^{\prime 2}\ = \ &b^{2}+2 \rho_{0} \rho_{M}+2 \rho_{0}^{2}+\sqrt{b^{4}-4 b^{2} \rho_{0}(\rho_{0}-\rho_{M})+4 \rho_{0}^{2}(\rho_{M}+\rho_{0})^{2}}, \\
M\ =\ &b^{2}+2 \rho_{0} \rho_{M}-2 \rho_{0}^{2}+\sqrt{b^{4}-4 b^{2} \rho_{0}(\rho_{0}-\rho_{M})+4 \rho_{0}^{2}(\rho_{M}+\rho_{0})^{2}}, \\
a\ =\ &b\left[1+\frac{4 \rho_{0}^{2}-b^{2}}{2 \rho_{0}(\rho_{M}-\rho_{0})+b^{2}+\sqrt{b^{4}-4 b^{2} \rho_{0}(\rho_{0}-\rho_{M})+4 \rho_{0}^{2}(\rho_{M}+\rho_{0})^{2}}}\right] ^{\frac{1}{2}}.
 \end{aligned}\label{metric14}
\end{eqnarray}
With these quantities,  all of coefficients in the metric (\ref{metric1}) can be expressed as  functions of the parameters $\rho_0$, $\rho_M $ and $\ b$.  With this functions, we can further study the shadow of the rotating squashed KK black hole (\ref{metric1}).

In the background of   the rotating squashed KK black hole (\ref{metric1}),
the Lagrange density of a photon propagation along null geodesics can be expressed as
\begin{eqnarray}
\begin{aligned}
\mathcal{L}=\frac{1}{2}g_{\mu \nu}\dot{x}^{\mu}\dot{x}^{\nu},
\end{aligned}\label{Hamiltonian}
\end{eqnarray}
where a dot represents a derivative with respect to affine parameter $\lambda$ along the geodesics.
Since all of metric functions $g_{\mu\nu}$ are independent of the coordinates $t$, $\phi$ and $\psi$,
there are three conserved quantities for the photon propagation
\begin{eqnarray}
\begin{aligned}
E&=-p_{t}=-g_{tt} \dot{t}-g_{t\phi} \dot{\phi}-g_{t \psi} \dot{\psi},\\
L_{\phi}&=p_{\phi}=g_{t\phi}\dot{t}+g_{\phi\phi}\dot{\phi}+g_{\phi \psi} \dot{\psi},\\
L_{\psi}&=p_{\psi}=g_{t\psi}\dot{t}+g_{\phi\psi}\dot{\phi}+g_{\psi\psi}\dot{\psi},
\end{aligned}\label{EL}
\end{eqnarray}
where $E$ is  energy of the photon. $L_{\phi}$ and $L_{\psi}$ denote its angular momentum in the $\phi$ and $\psi$ directions, respectively. With these conserved quantities, the null geodesics for a photon can be further simplified as
\begin{eqnarray}
\begin{aligned}
&\dot{t}=\frac{D(\rho)E-H(\rho)L_{\psi}}{A(\rho)D(\rho)+H(\rho)^{2}},\quad\quad\quad\quad \dot{\phi}=\frac{L_{\phi}-\cos{\theta} L_{\psi}}{\sin^{2}{\theta} \ C(\rho)},\\
&\dot{\psi}=\frac{H(\rho)E+A(\rho)L_{\psi}}{A(\rho)D(\rho)+H(\rho)^{2}}
-\frac{(L_{\phi}-L_{\psi}\cos{\theta})\cos{\theta}}{\sin^{2}{\theta} \ C(\rho)},
\end{aligned}\label{geodesics0}
\end{eqnarray}
and
\begin{eqnarray}
\begin{aligned}
&C(\rho)^2\ \dot{\rho}^{2}=R(\rho)=\frac{C(\rho)^2}{B(\rho)}\bigg[\frac{D(\rho)E^{2}
-2H(\rho)EL_{\psi}-A(\rho)L_{\psi}^{2}}{A(\rho)D(\rho)+H(\rho)^{2}}-\frac{L_{\phi}^{2}+Q}{C(\rho)}\bigg],\\
&C(\rho)^{2}\ \dot{\theta}^{2}=\Theta(\theta)=-\frac{(L_{\phi}-\cos{\theta} L_{\psi})^2}{\sin^{2}{\theta}}+L_{\phi}^{2}+Q,
\end{aligned}\label{geodesics}
\end{eqnarray}
where $Q$ is the Carter constant.  From  the  above equation, we find that the $\theta-$ component  equation is independent of the rotation parameter and it is exactly
identical to the equation in the static squashed KK black hole
spacetime, which would yield some special properties of the shadow of the black hole (\ref{metric1}).
The circular orbits satisfy
\begin{eqnarray}
\begin{aligned}
R(\rho)=0,\qquad \qquad  R'(\rho)=0.
\end{aligned}\label{circular orbit0}
\end{eqnarray}
It is well known that the unstable circular orbits are very important to determine the boundary of the shadow casted by a black hole.

\section{Shadow of a rotating squashed Kaluza-Klein black hole}

We assume that the observer is located in the spatial infinity in  the rotating squashed KK black hole spacetime. As $\rho\rightarrow\infty$, the metric (\ref{metric1}) becomes
\begin{eqnarray}
d s^{2}&=&-d t^{2}+ d \rho^{2}+\rho^2(d \theta^{2}+\sin ^{2} \theta d \phi^{2})+\frac{r'^2_{\infty}}{4}(d \psi+\cos \theta d \phi)^{2},
\end{eqnarray}
which describes a four-dimensional Minkowski spacetime with a constant twisted
$S_1$ fiber. Setting $dw=\frac{r'_{\infty}}{2}(d \psi+\cos \theta d \phi)$, and making the coordinate transformation
\begin{eqnarray}\label{zuob1}
x=\rho\sin\theta\cos\phi,\quad\quad\quad\quad y=\rho\cos\theta, \quad\quad\quad\quad z=\rho\sin\theta\sin\phi,
\end{eqnarray}
one can find that the above metric can be rewritten as
\begin{eqnarray}
d s^{2}&=&-d t^{2}+d x^{2}+dy^2+dz^2+ dw^{2},
\end{eqnarray}
which has a form of five-dimensional Minkowski metric. Thus, in the rotating squashed KK black hole spactime, the observer basis at the spatial infinity $\left\{e_{\hat{t}}, e_{\hat{x}}, e_{\hat{y}}, e_{\hat{z}}, e_{\hat{w}}\right\}$  can be expanded as a form in the coordinate basis $\left\{\partial_{t}, \partial_{\rho}, \partial_{\theta}, \partial_{\phi}, \partial_{\psi}\right\}$ \cite{w31,w32}
\begin{eqnarray}
e_{\hat{\mu}}=e_{\hat{\mu}}^{\nu} \partial_{\nu}.\label{coordinate1}
\end{eqnarray}
Here $e^{\nu}_{\hat{\mu}}$ is the transform matrix which satisfies $g_{\mu\nu}e^{\mu}_{\hat{\alpha}}e^{\nu}_{\hat{\beta}}
=\eta_{\hat{\alpha}\hat{\beta}}$. $\eta_{\hat{\alpha}\hat{\beta}}$ is the five-dimensional Minkowski metric.
In general, it is not unique for the transformation (\ref{coordinate1}) which satisfies both the
spatial rotations and Lorentz boosts. For a rotating squashed KK black hole spacetime (\ref{metric1}), one can choice a convenient decomposition associated with a reference
frame with zero axial angular momentum in relation to spatial infinity \cite{w31,w32}
\begin{eqnarray}
e_{\hat{\mu}}^{\nu}=\left( \begin{array}{ccccc}
{\zeta} & {0} & {0} & {\lambda} & {\chi} \\
 {0} & {A^{\rho}} & {0} & {0} & {0} \\
  {0} & {0} & {A^{\theta}} & {0} & {0} \\
   {0} & {0} & {0} & {A^{\phi}} & {\varepsilon} \\
    {0} & {0} & {0} & {0} & {A^{\psi}}\end{array}\right),\label{coordinate3}
\end{eqnarray}
where $\zeta$,  $\lambda$, $\chi$, $\varepsilon$, $A^r$, $A^{\theta}$,  $A^{\phi}$  and $\ A^{\psi}$ are real coefficients. According to the Minkowski normalization
\begin{eqnarray}
e_{\hat{\mu}}e^{\hat{\nu}}=\delta_{\hat{\mu}}^{\;\hat{\nu}},
\end{eqnarray}
one can obtain
\begin{eqnarray}
\begin{aligned}
&A^{\rho}=\frac{1}{\sqrt{g_{\rho\rho}}},\quad \quad \quad A^{\theta}=\frac{1}{\sqrt{g_{\theta \theta}}},\quad\quad\quad  A^{\phi}=\sqrt{\frac{g_{\psi \psi}}{g_{\phi \phi}g_{\psi \psi}-g^2_{\phi \psi}}},\\
&A^{\psi}=\frac{1}{\sqrt{g_{\psi \psi}}},\quad\quad\quad \varepsilon=-\frac{g_{\phi \psi}}{g_{\psi \psi}}\sqrt{\frac{g_{\psi \psi}}{g_{\phi \phi}g_{\psi \psi}-g^2_{\phi \psi}}},\\
&\zeta=\frac{\sqrt{g_{\phi \psi}^{2}-g_{\phi \phi} g_{\psi \psi}}}{\sqrt{g_{t t} g_{\phi \phi} g_{\psi \psi}+2 g_{t \phi} g_{t \psi} g_{\phi \psi}-g_{t \phi}^{2} g_{\psi \psi}-g_{t \psi}^{2} g_{\phi \phi}-g_{\phi \psi}^{2} g_{t t}}},\\
&\lambda=\frac{g_{t\phi} g_{\psi \psi}-g_{t\psi} g_{\phi \psi}}{\sqrt{g_{\phi \psi}^{2}-g_{\phi \phi} g_{\psi \psi}} \sqrt{g_{t t} g_{\phi \phi} g_{\psi \psi}+2 g_{t \phi} g_{t \psi} g_{\phi \psi}-g_{t \phi}^{2} g_{\psi \psi}-g_{t \psi}^{2} g_{\phi \phi}-g_{\phi \psi}^{2} g_{t t}}},\\
&\chi=\frac{g_{t\psi} g_{\phi \phi}-g_{t\phi} g_{\phi \psi}}{\sqrt{g_{\phi \psi}^{2}-g_{\phi \phi} g_{\psi \psi}} \sqrt{g_{t t} g_{\phi \phi} g_{\psi \psi}+2 g_{t \phi} g_{t \psi} g_{\phi \psi}-g_{t \phi}^{2} g_{\psi \psi}-g_{t \psi}^{2} g_{\phi \phi}-g_{\phi \psi}^{2} g_{t t}}}.
\end{aligned}\label{coordinate4}
\end{eqnarray}
Therefore,  the locally measured five-momentum $p^{\hat{\mu}}$ of a photon is computed by the projection of its five-momentum $p^{\mu}$  into $e_{\hat{\mu}}$
\begin{eqnarray}
\begin{aligned} p^{\hat{t}} &=\zeta E-\lambda p_{\phi}-\chi p_{\psi}, \quad\quad\quad\quad
p^{\hat{x}} =\frac{1}{\sqrt{g_{\rho \rho}}}p_{\rho}, \\
p^{\hat{y}} &=\frac{1}{\sqrt{g_{\theta \theta}}}p_{\theta},\quad\quad\quad\quad
p^{\hat{z}} =\sqrt{\frac{g_{\psi \psi}}{g_{\phi \phi}g_{\psi \psi}-g^2_{\phi \psi}}}\bigg(p_{\phi}-\frac{g_{\phi \psi}}{g_{\psi \psi}}p_{\psi}\bigg) ,\quad\quad\quad\quad
p^{\hat{w}} =\frac{1}{\sqrt{g_{\psi \psi}}} p_{\psi}.
\end{aligned}\label{coordinate5}
\end{eqnarray}
The four-vector $\vec{p}$ is the photon's linear momentum with components $\ p_{\hat{x}}$, $p_{\hat{y}} $,  $p_{\hat{z}} $  and $p_{\hat{w}}$ in the orthonormal basis $\left\{e_{\hat{x}}, e_{\hat{y}}, e_{\hat{z}}, e_{\hat{w}}\right\}$,
\begin{eqnarray}
\begin{aligned}
\vec{p}&=p^{\hat{x}} e_{\hat{x}}+p^{\hat{y}} e_{\hat{y}}+p^{\hat{z}}e_{\hat{z}}+p^{\hat{w}} e_{\hat{w}}.
\end{aligned}
\end{eqnarray}
Combing the transformation (\ref{zuob1}) with the geometry of the photon's linear momentum, we have
\begin{eqnarray}
p^{\hat{x}} =\tilde{p} \cos \alpha \cos \beta, \quad\quad\quad
p^{\hat{y}} =\tilde{p} \sin \alpha,\quad\quad\quad
p^{\hat{z}} =\tilde{p} \cos \alpha \sin \beta,\quad\quad\quad
p^{\hat{w}}=\sqrt{|\vec{p}|^2-\tilde{p}^2}.\label{coordinate6}
\end{eqnarray}
where $\tilde{p}=\sqrt{(p^{\hat{x}})^2+(p^{\hat{y}})^2+(p^{\hat{z}})^2}$. In general, there should be three  celestial coordinates of photon's image in the observer's sky in the five-dimensional background spacetime. Since in the spatial infinity the rotating squashed KK black hole spacetime has a structure of four-dimensional Minkowski spacetime with a constant twisted $S_1$ fiber, the fifth dimension is compacted for the observer in the position far from black hole in this case. This allows us to adopt the usual celestial coordinates $(x,y)$ in the four-dimensional case to describe the position of  photon's image in the observer's sky because the celestial coordinates $(x,y)$ could be observed really by our astronomical
experiments. Moreover, it is very convenient for us to compare our results with those obtained in the
usual four-dimensional black hole spacetimes. The celestial coordinates $(x,y)$ for the  photon's image in observer's sky for a rotating squashed KK black hole (\ref{metric1}) are
\begin{eqnarray}
x &=&-\lim _{\rho_{obs} \rightarrow \infty} \rho_{obs}\tan\beta=-\lim _{\rho_{obs} \rightarrow \infty}\rho_{obs} \frac{p^{\hat{z}}}{p^{\hat{x}}}=-\frac{\xi_{\phi}-\xi_{\psi}\cos\theta_0}{\sin{\theta_0}}\frac{r'_{\infty}}
{\sqrt{r'^2_{\infty}-4\xi^2_{\psi}}},\nonumber\\
y &=&\lim _{\rho_{obs} \rightarrow \infty} \rho_{obs}\frac{\tan\alpha}{\cos{\beta}}=\lim _{\rho_{obs} \rightarrow \infty} \rho_{obs} \frac{p^{\hat{y}}}{p^{\hat{x}}}= \frac{r'_{\infty}}{\sqrt{r'^2_{\infty}-4\xi^2_{\psi}}} \sqrt{\eta+\xi_{\phi}^2-\bigg(\frac{\xi_{\phi}-\xi_{\psi}\cos\theta_0}{\sin{\theta_0}}\bigg)^2}.
\label{coordinate7}
\end{eqnarray}
Here $\xi_{\phi}\equiv L_{\phi}/E$, $\xi_{\psi}\equiv L_{\psi}/E$, $\eta\equiv Q/E^2$, and $\rho_{obs}$ is the distance between black hole and the observer. $\theta_{0}$ is  the inclination angle of observer. From Eq.(\ref{coordinate7}), it is easy to obtain
\begin{eqnarray}
x^{2}+y^{2}=\frac{r'^2_{\infty}(\eta+\xi_{\phi}^2)}{r'^2_{\infty}-4\xi^2_{\psi}},\label{coordinate9}
\end{eqnarray}
which implies that the shape of the shadow is a perfect black disk for a rotating squashed KK black hole (\ref{metric1}). It is different from that of a usual rotating black hole. This  special property of shadow of a rotating squashed KK black hole could be explained by a fact that the $\theta$- component  equation is independent of the rotation parameter in this special black hole spacetime.
Moreover, we find that the radius of the image in the observer's sky caused by the photon with $\xi_{\psi}$  is
\begin{eqnarray}
R_s=\frac{r'_{\infty}\sqrt{\eta+\xi_{\phi}^2}}{\sqrt{r'^2_{\infty}-4\xi^2_{\psi}}}
=\frac{r'_{\infty}}{\sqrt{r'^2_{\infty}-4\xi^2_{\psi}}}\sqrt{\frac{C(\rho_{ps})[D(\rho_{ps})
-2H(\rho_{ps})\xi_{\psi}-A(\rho_{ps})\xi^2_{\psi}]}
{A(\rho_{ps})D(\rho_{ps})+H(\rho_{ps})^{2}}},\label{coordinate10}
\end{eqnarray}
\begin{figure}[ht]
\begin{center}
\subfigure[]{\includegraphics[width=4cm]{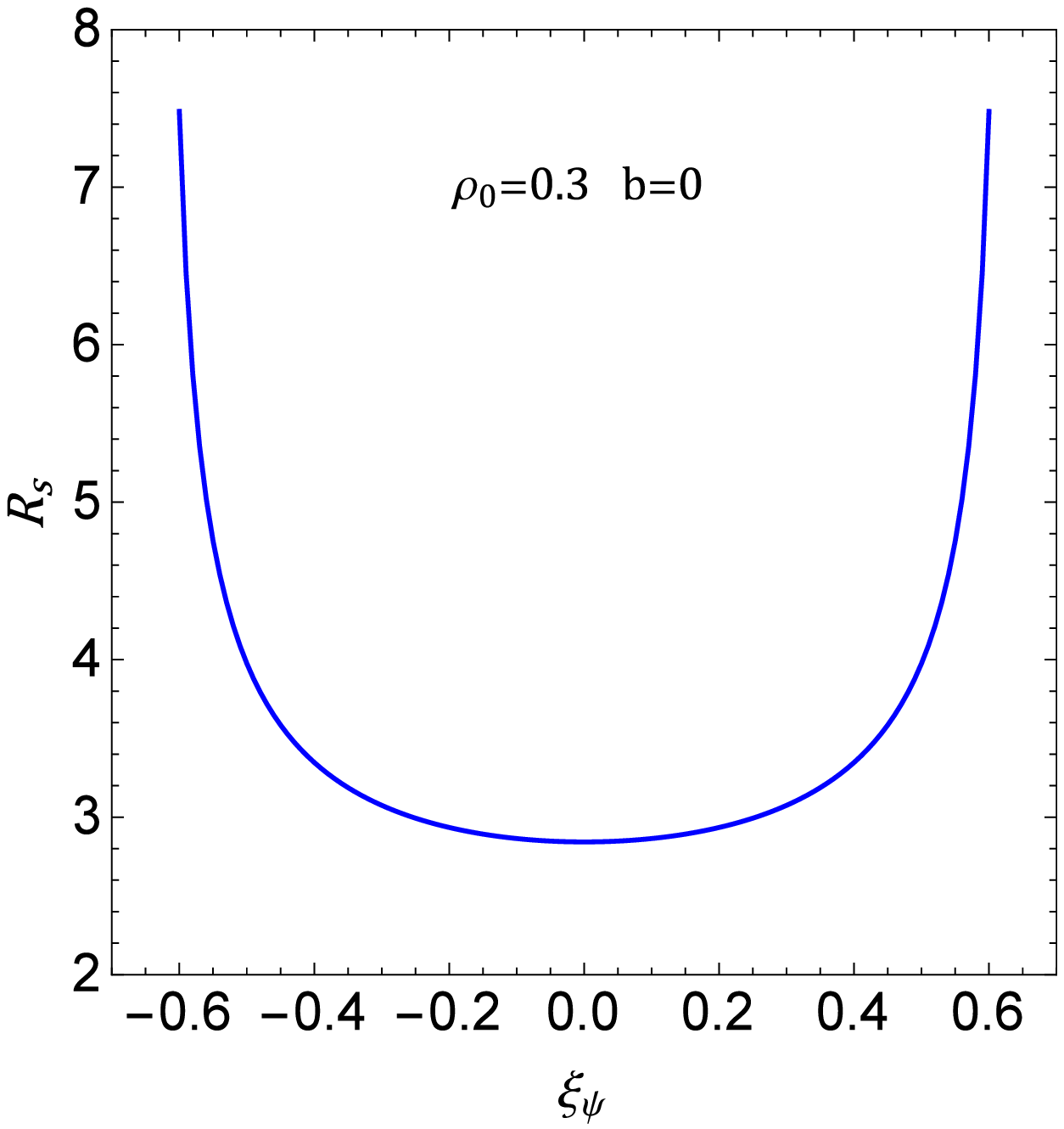}}\;
\subfigure[]{\includegraphics[width=4cm]{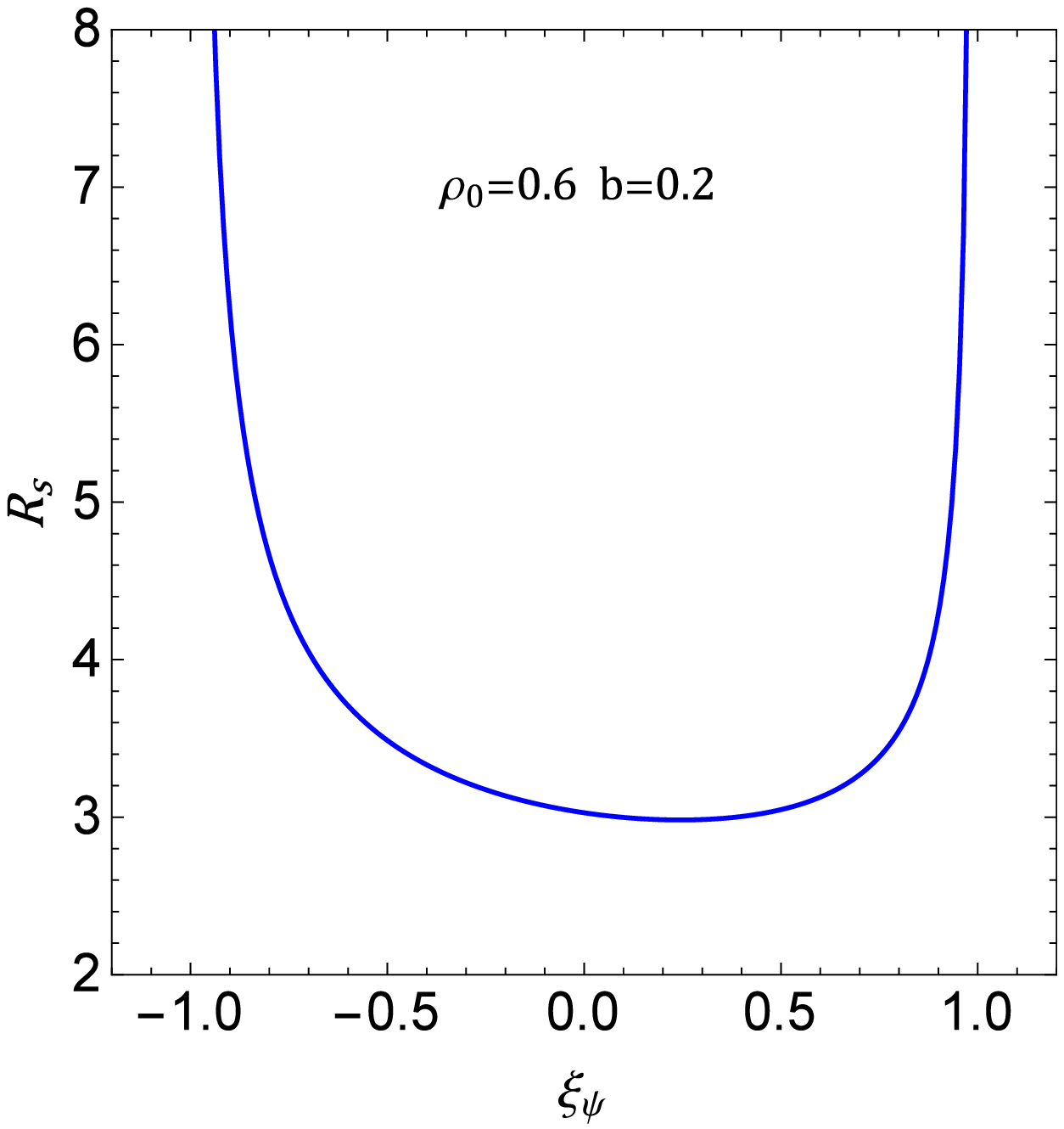}}
\subfigure[]{\includegraphics[width=4cm]{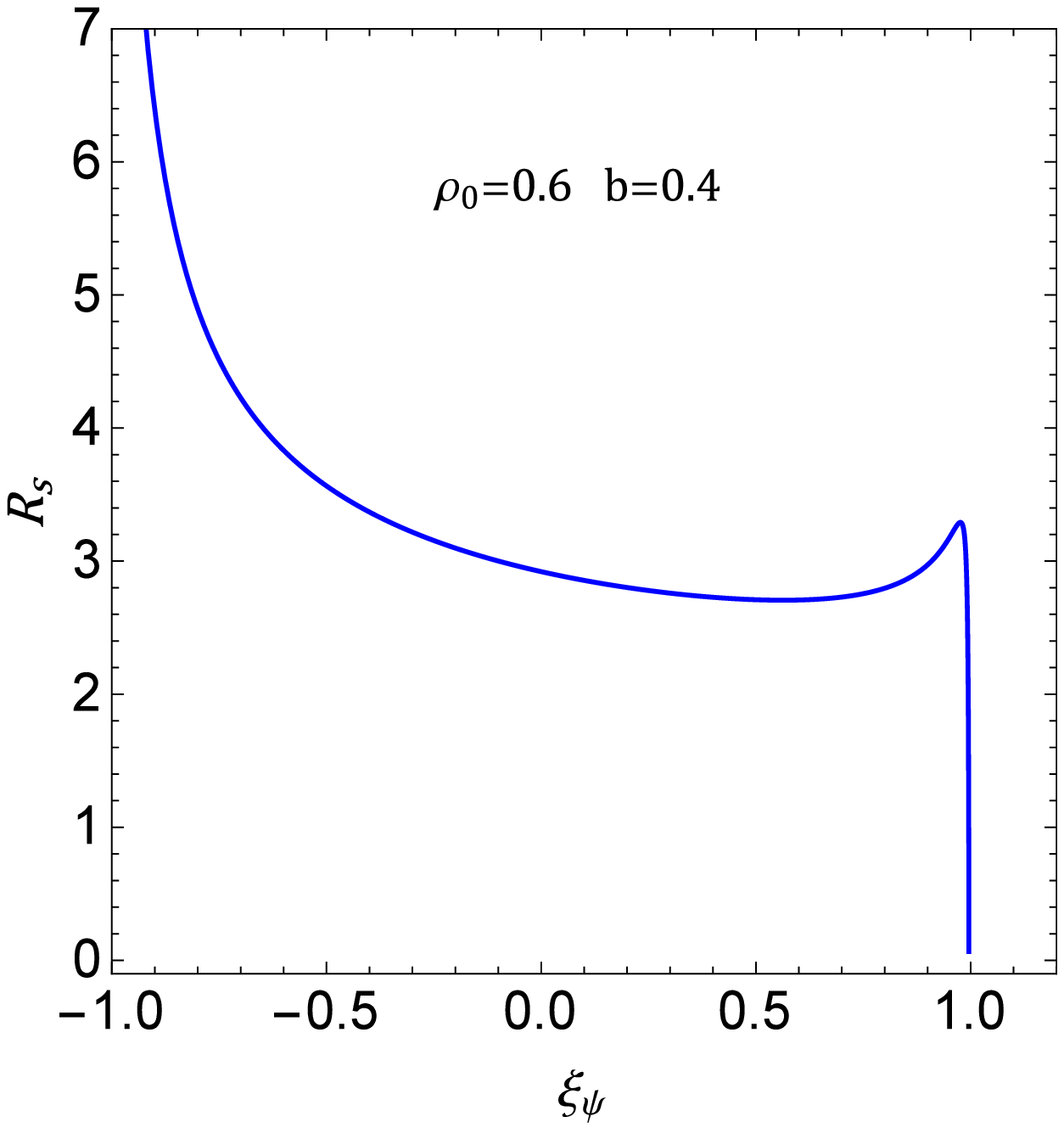}}
\subfigure[]{\includegraphics[width=4cm]{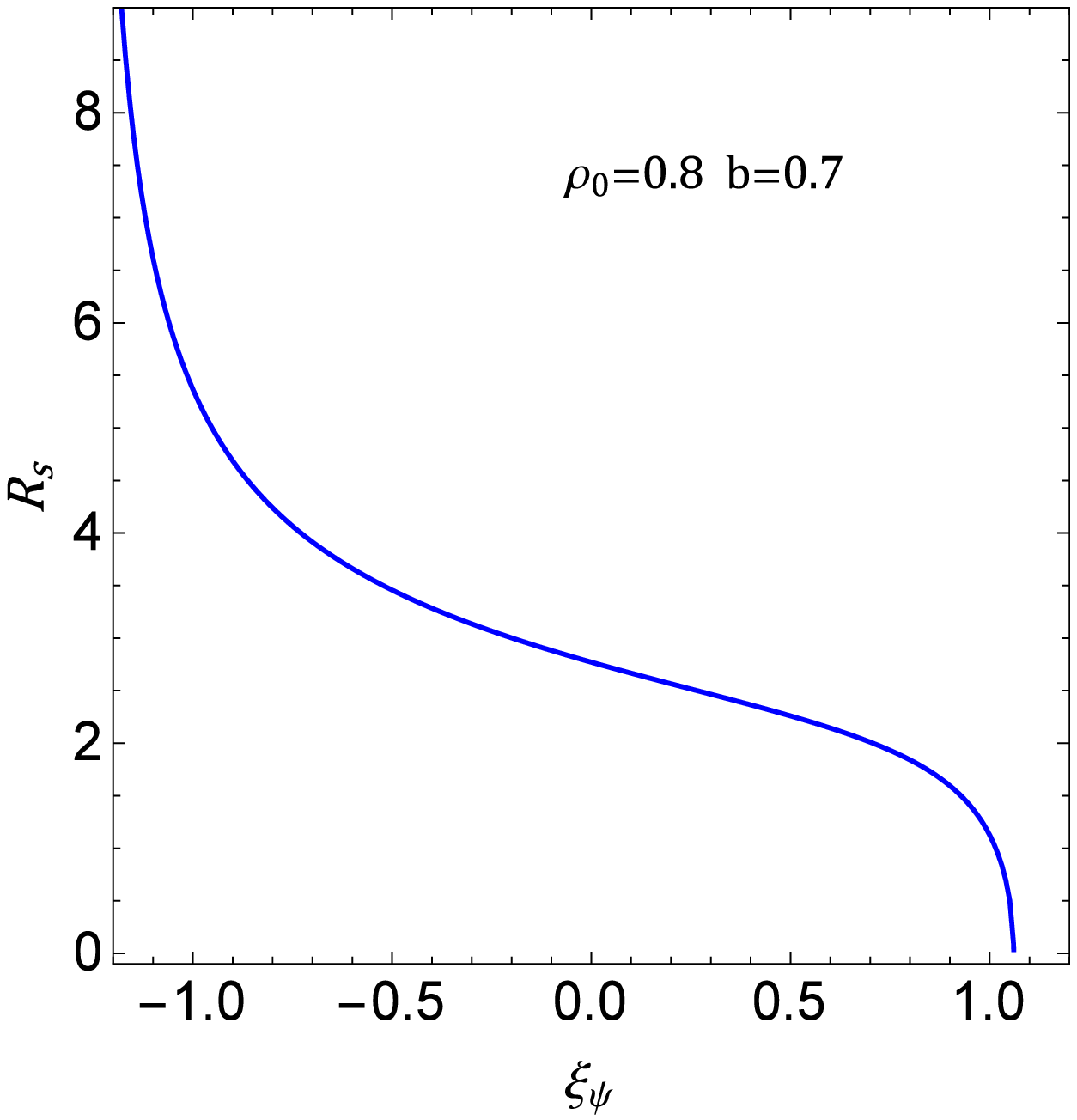}}
\caption{Changes of $R_s$ with $\xi_{\psi}$  for the rotating squashed KK black hole with different $\rho_0$ and $b$. Here we set $\rho_M=1$. }\label{fig1}
\end{center}
\end{figure}
where $\rho_{ps}$ is the photon sphere radius of photon with the specific angular momentum $\xi_{\psi}$.
Obviously, the radius $R_s$ is a function of parameters $\xi_{\psi}$, $\rho_0$, $\rho_M$ and $b$. Thus, for the black hole with fixed parameters $\rho_0$, $\rho_M$ and $b$, the quantity $R_s$ has different value for the photon with different $\xi_{\psi}$. This means that the radius of the black hole shadow is determined by the minimum value of $R_s$, i.e.,
\begin{eqnarray}
R_{BH}|_{(\rho_0,\rho_M, b)}=\text{Minimum} [R_s(\rho_0, \rho_M, b,\xi_{\psi})]|_{(\rho_0,\rho_M, b)}.\label{coordinate10bh}
\end{eqnarray}
\begin{figure}
\begin{center}
\subfigure[$\xi_{\psi}=0.9$]{\includegraphics[width=4cm]{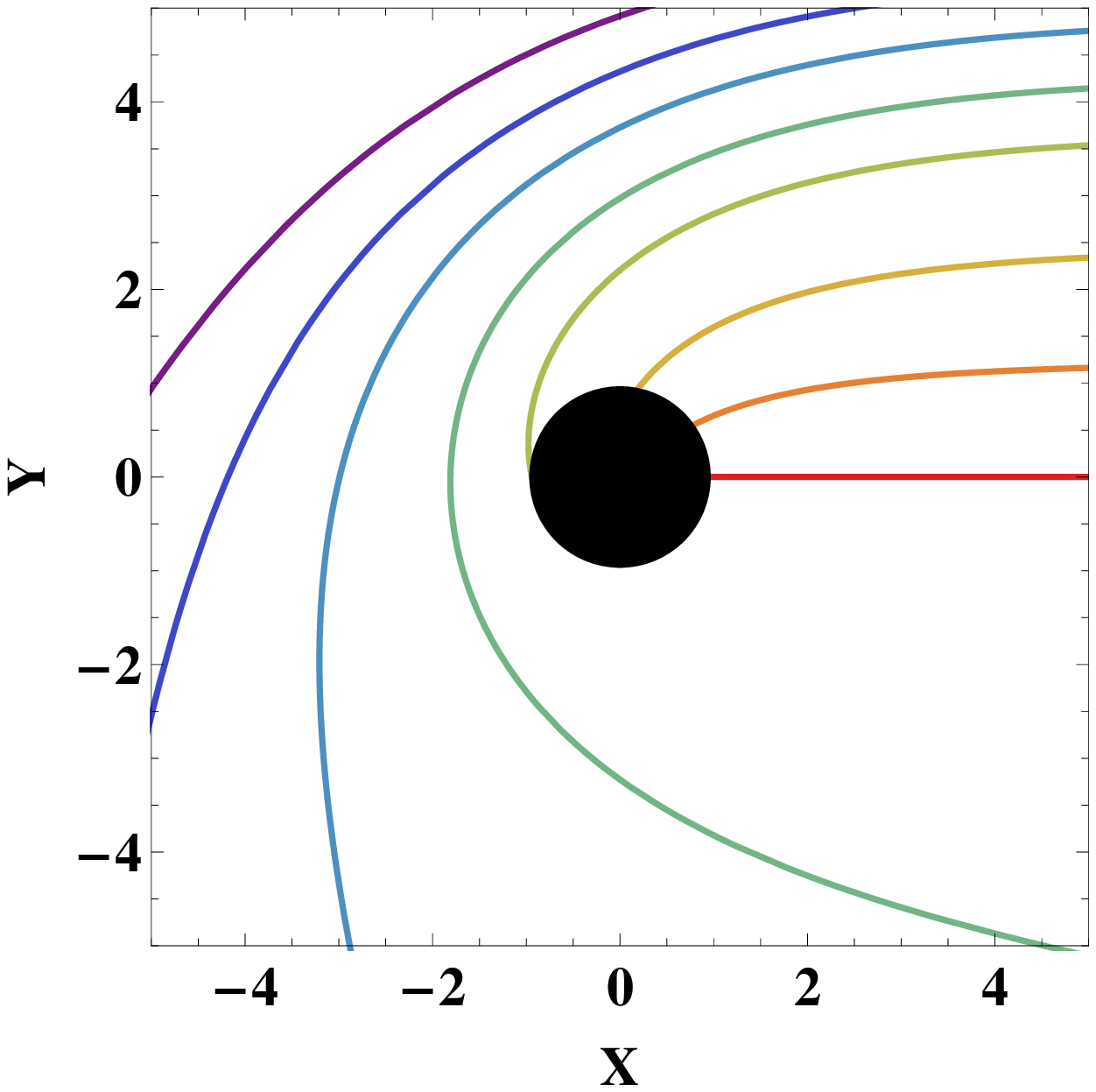}}\;
\subfigure[$\xi_{\psi}=0.94$]{\includegraphics[width=4cm]{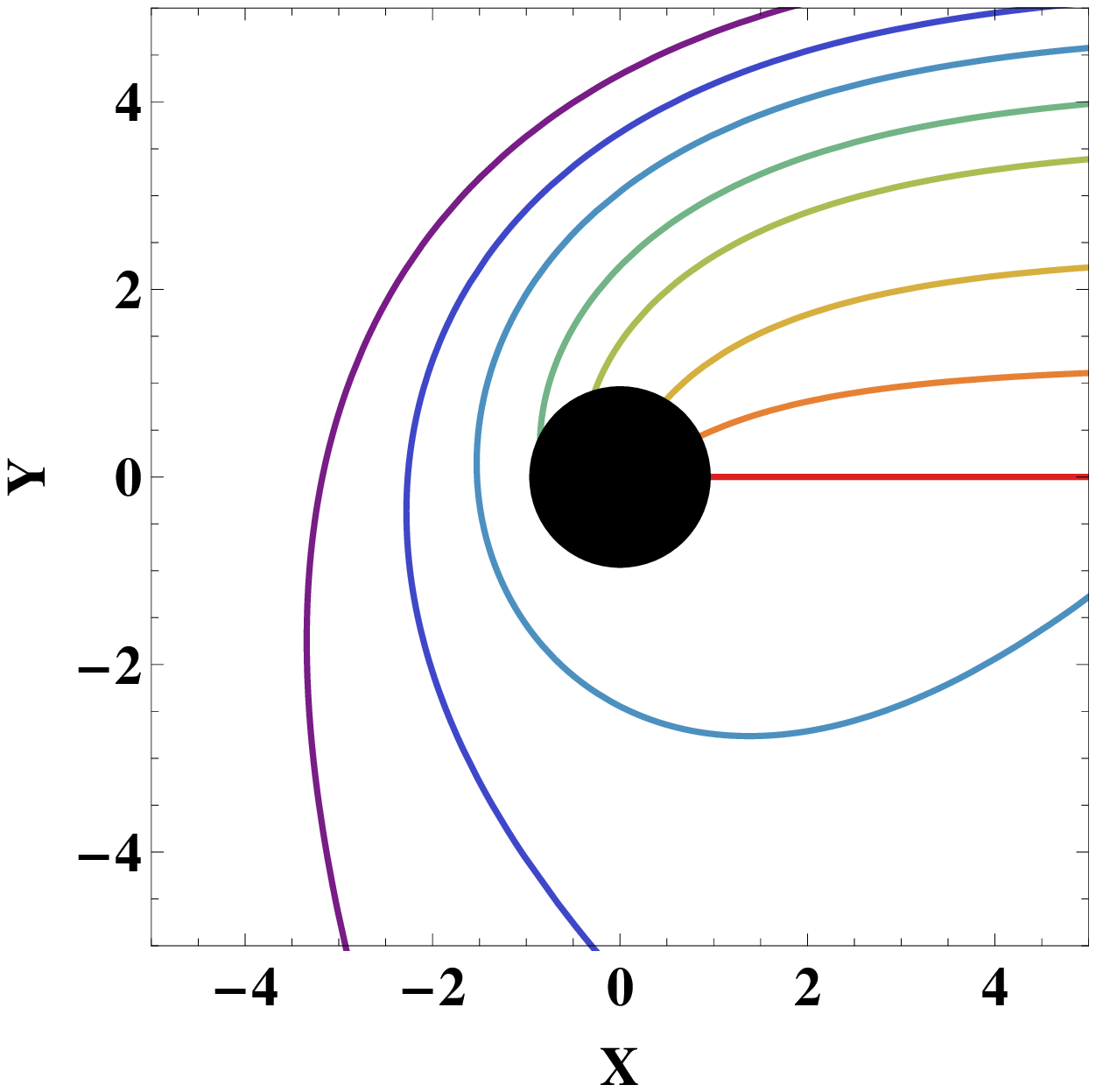}}
\subfigure[$\xi_{\psi}=0.95$]{\includegraphics[width=4cm]{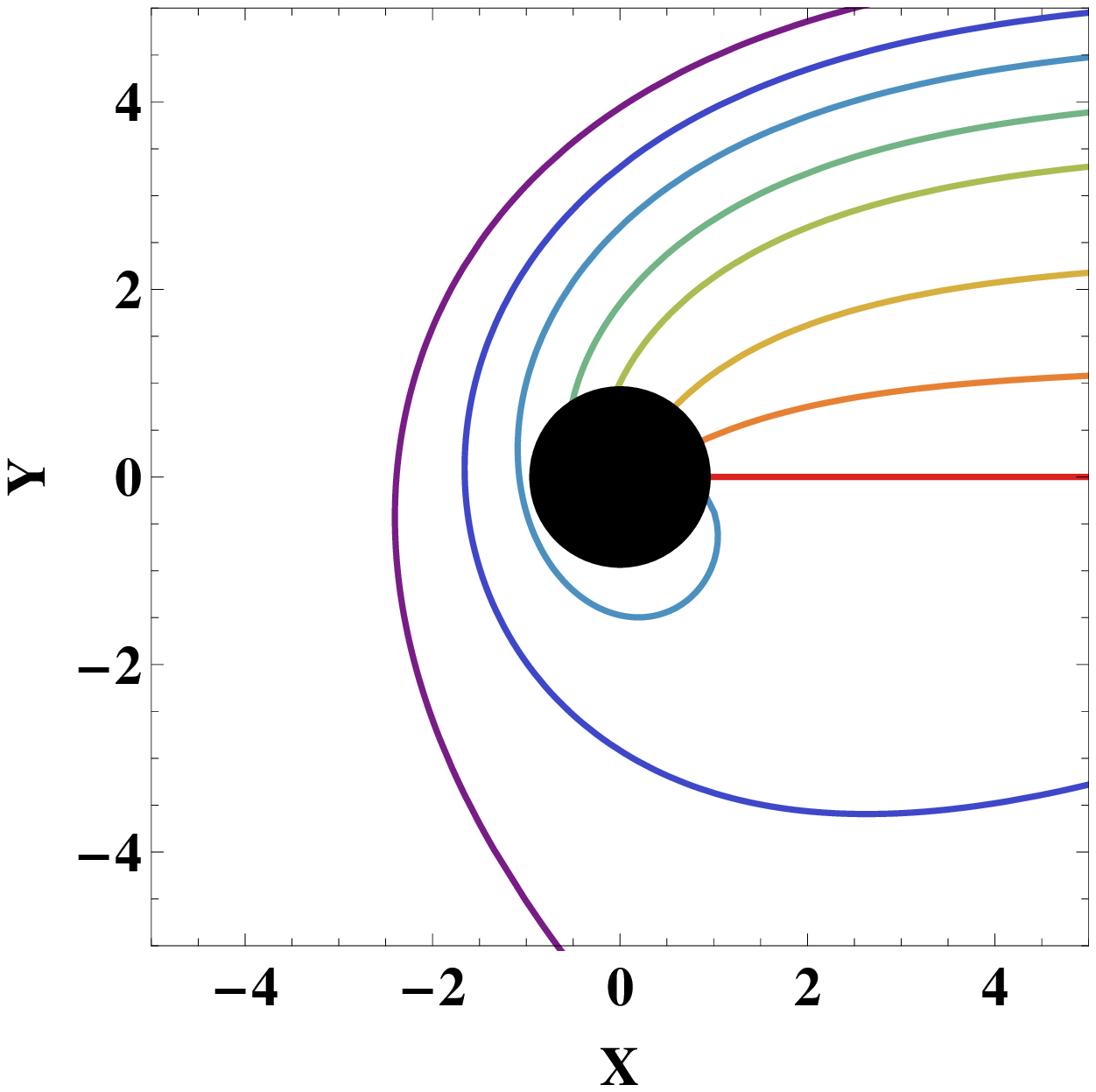}}
\subfigure[$\xi_{\psi}=0.96$]{\includegraphics[width=4cm]{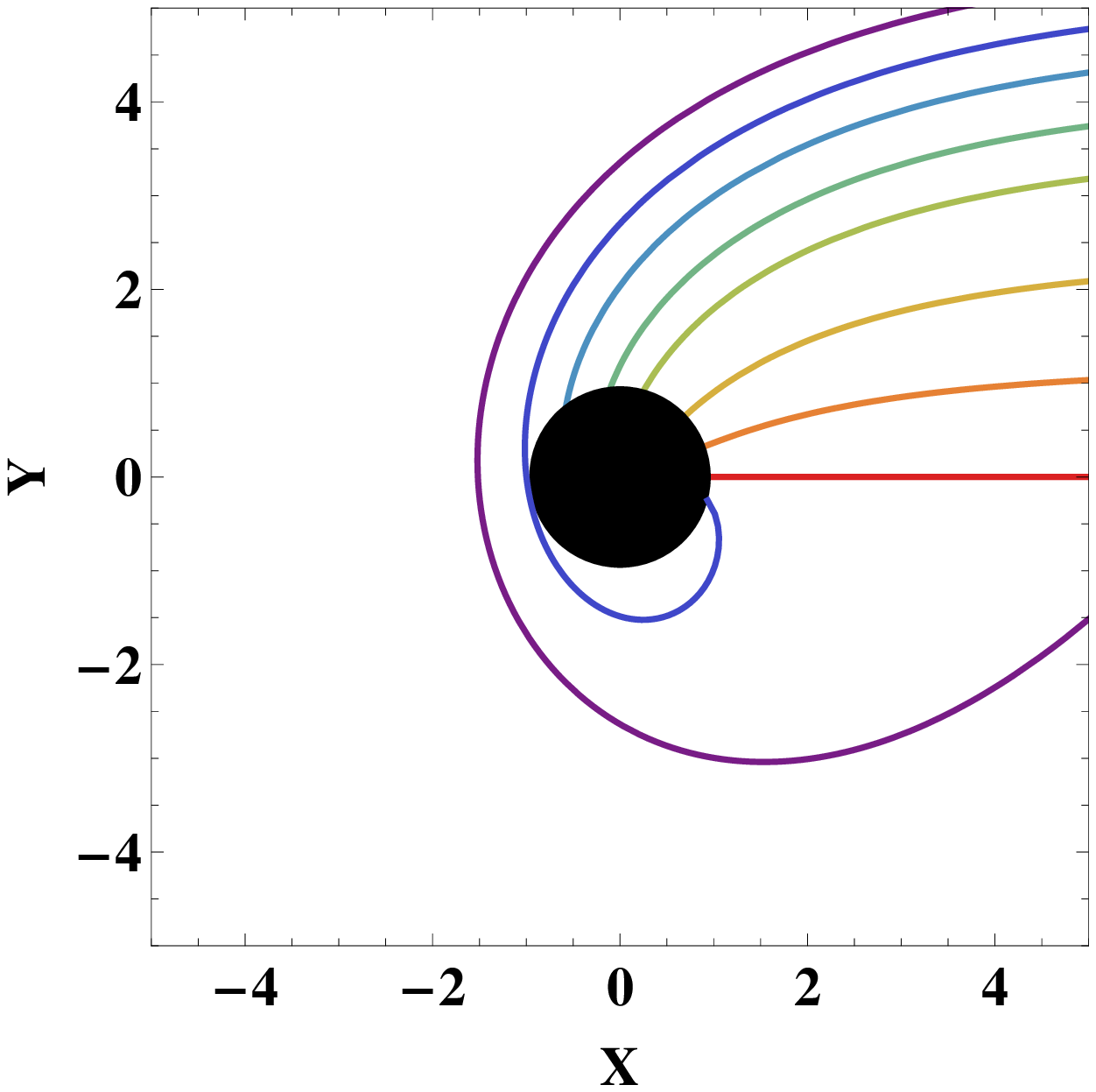}}
\caption{Propagation of photon with different $\xi_{\psi}$ in  the rotating squashed KK black hole spacetime with the fixed $\rho_0=0.6$ and $b=0.2$. Here we set $\rho_M=1$, $X=\rho \sin\theta\cos\phi$ and $Y=\rho (\sin\theta\cos\phi+\cos\theta)/\sqrt{2}$. }\label{fig1s1}
\end{center}
\end{figure}
Obviously, it is heavily influenced by the specific angular momentum $\xi_{\psi}$ of the photon. From Eq.(\ref{coordinate10}), one can find that $\xi_{\psi}$ should be limited in the range $\xi_{\psi}\in(-\frac{r'_{\infty}}{2},\frac{r'_{\infty}}{2})$. Here, in Fig.\ref{fig1}, we present the changes of
$R_s$ with $\xi_{\psi}$ for different black hole parameters in the rotating squashed KK black hole spacetime, which indicates that the specific angular momentum of photon $\xi_{\psi}$ affects sharply the black hole shadow. For the fixed $b=0$, $\rho_0=0.3$, we find that $R_s$ has a minimum value $R_{s_{\text{min}}}=2.8426$ as $\xi_{\psi}=0$, which means that there exists black hole shadow with its radius $R_{BH}=2.8426$ in this case. However, for the case with $b=0.4$ and $\rho_0=0.6$, we obtain the minimum value $R_{s_{\text{min}}}=0$ corresponding $\xi_{\psi}=0.996112$, which is in the range $\xi_{\psi}\in(-\frac{r'_{\infty}}{2},\frac{r'_{\infty}}{2})$. The similar behavior also appears in the case with $b=0.7$ and $\rho_0=0.8$. The special behavior $R_{s_{\text{min}}}=0$  means that there is no black shadow for a black hole in these cases, which is novel since it is impossible to appear in the usual black hole spacetimes.
\begin{figure}
\begin{center}
\subfigure[$\xi_{\psi}=0.9$]{\includegraphics[width=4cm]{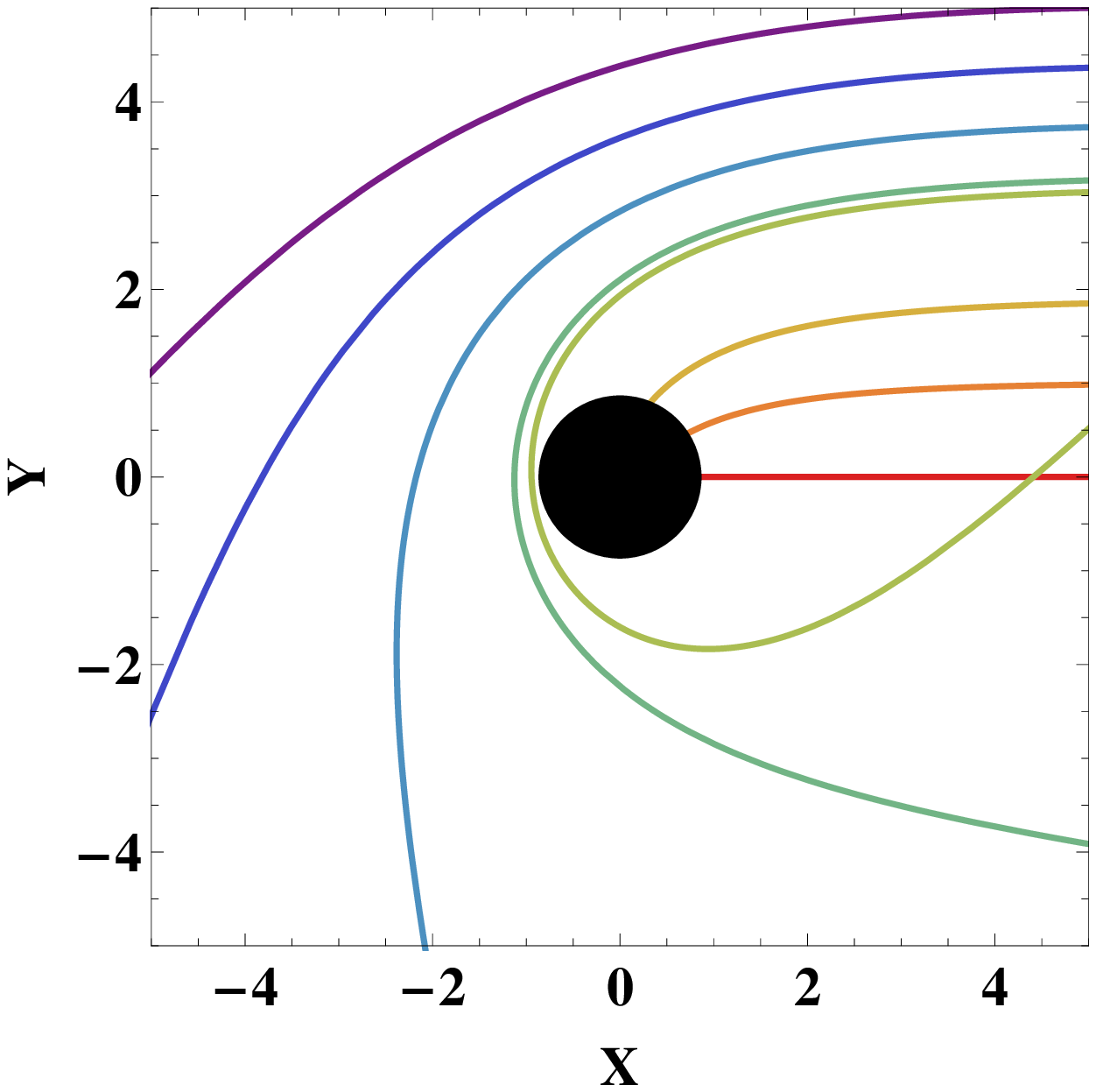}}\;
\subfigure[$\xi_{\psi}=0.95$]{\includegraphics[width=4cm]{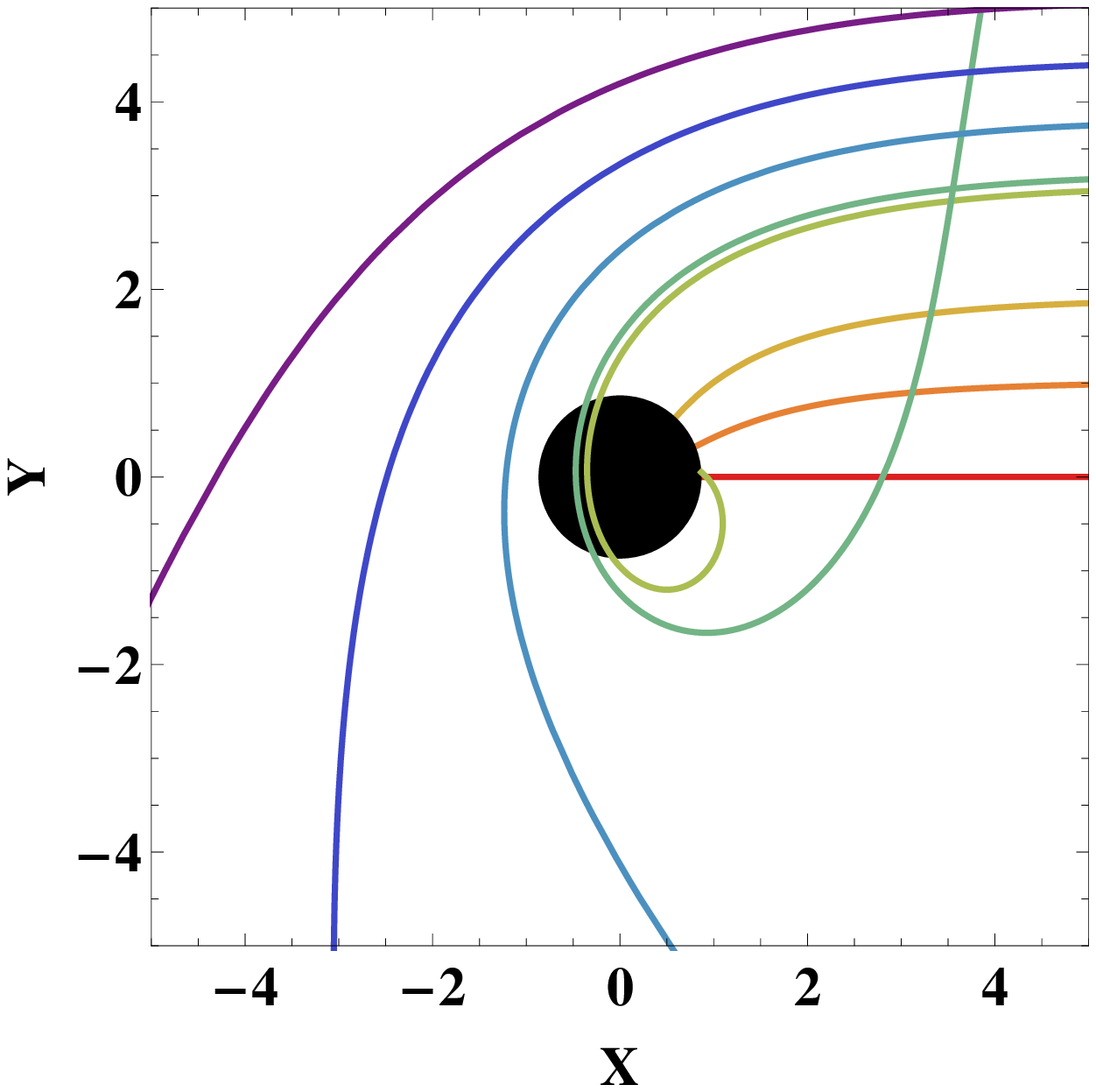}}
\subfigure[$\xi_{\psi}=0.97$]{\includegraphics[width=4cm]{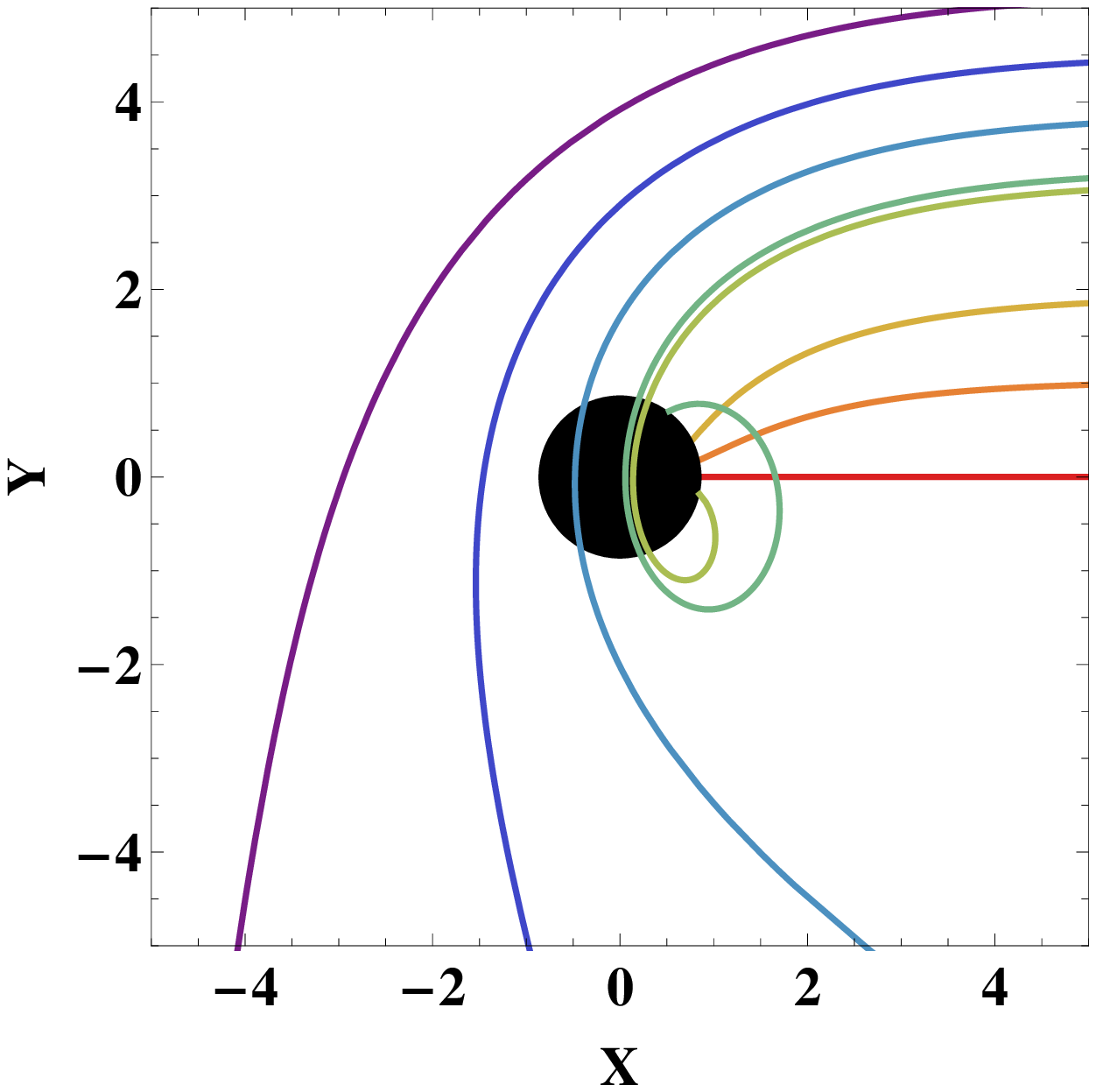}}
\subfigure[$\xi_{\psi}=0.99$]{\includegraphics[width=4cm]{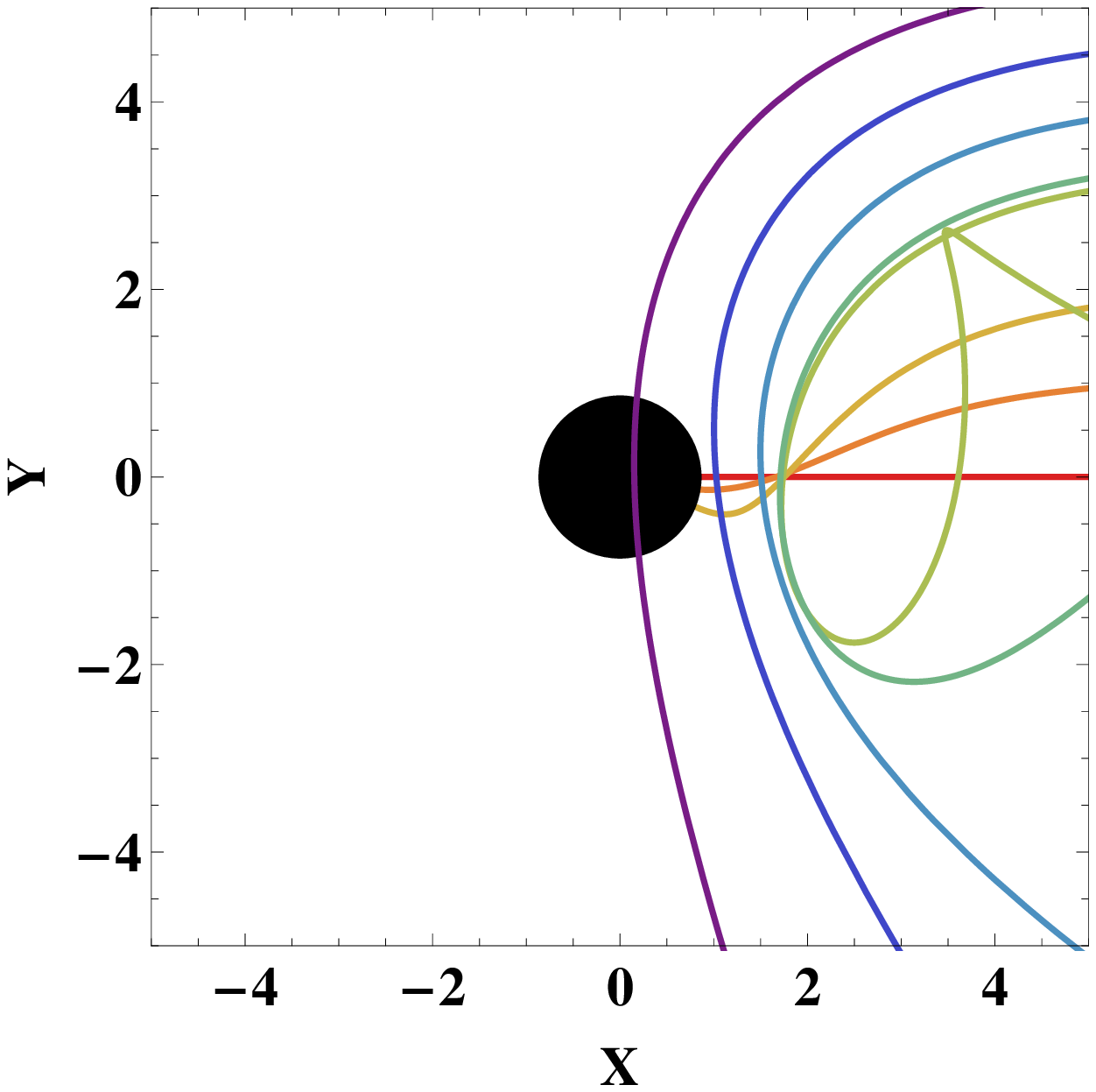}}\\
\subfigure[$\xi_{\psi}=0.992$]{\includegraphics[width=4cm]{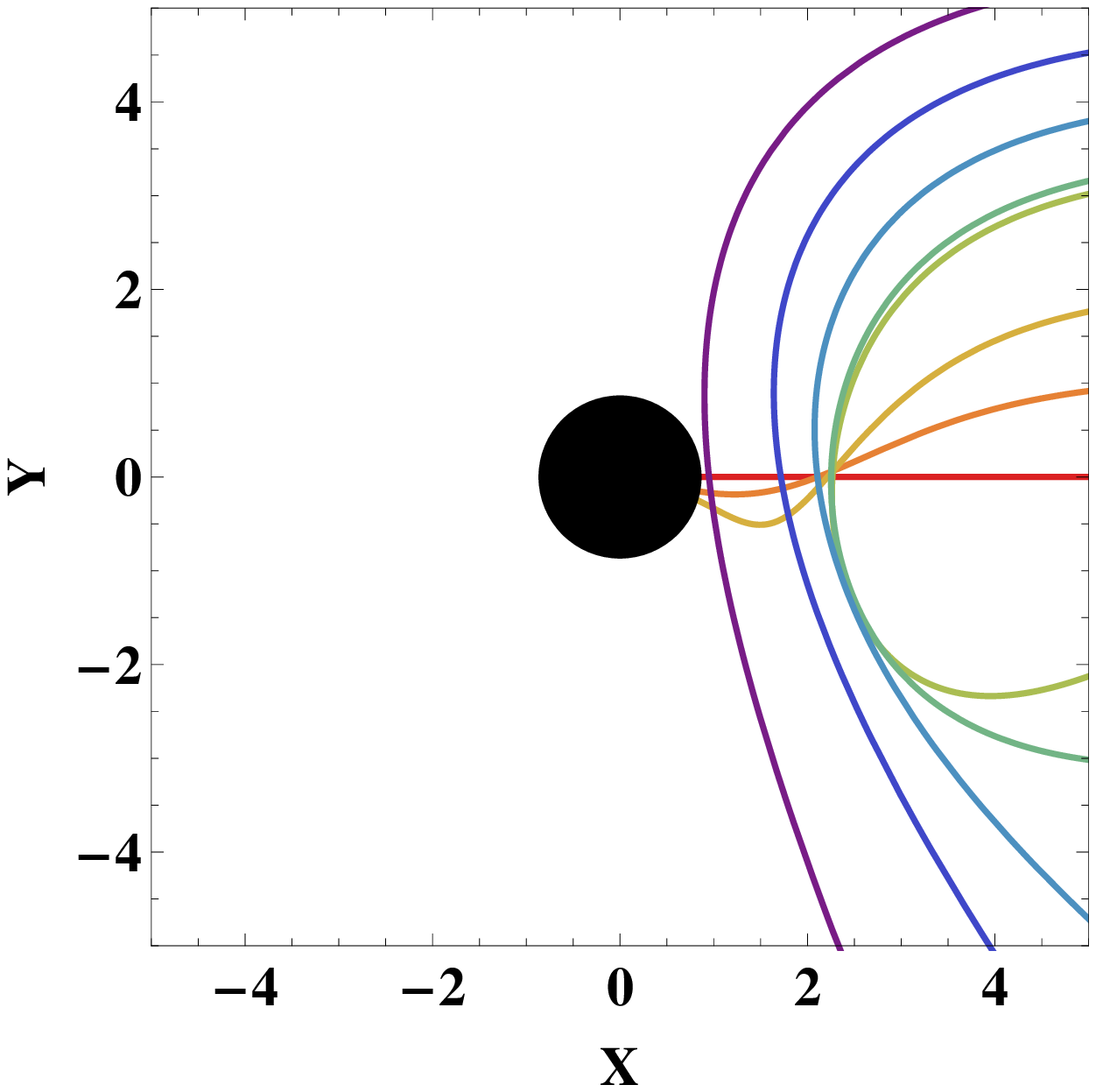}}\;
\subfigure[$\xi_{\psi}=0.994$]{\includegraphics[width=4cm]{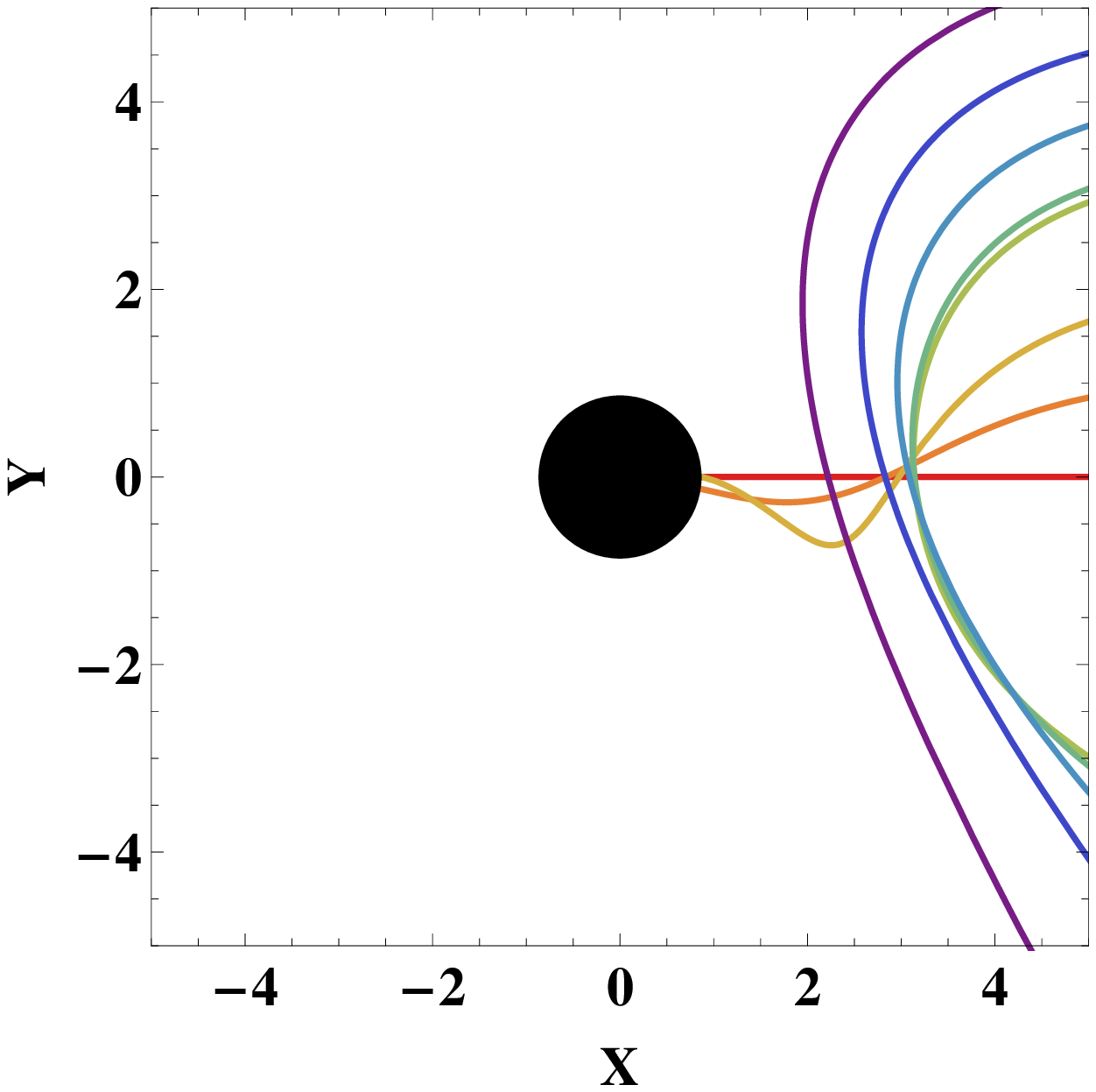}}
\subfigure[$\xi_{\psi}=0.996$]{\includegraphics[width=4cm]{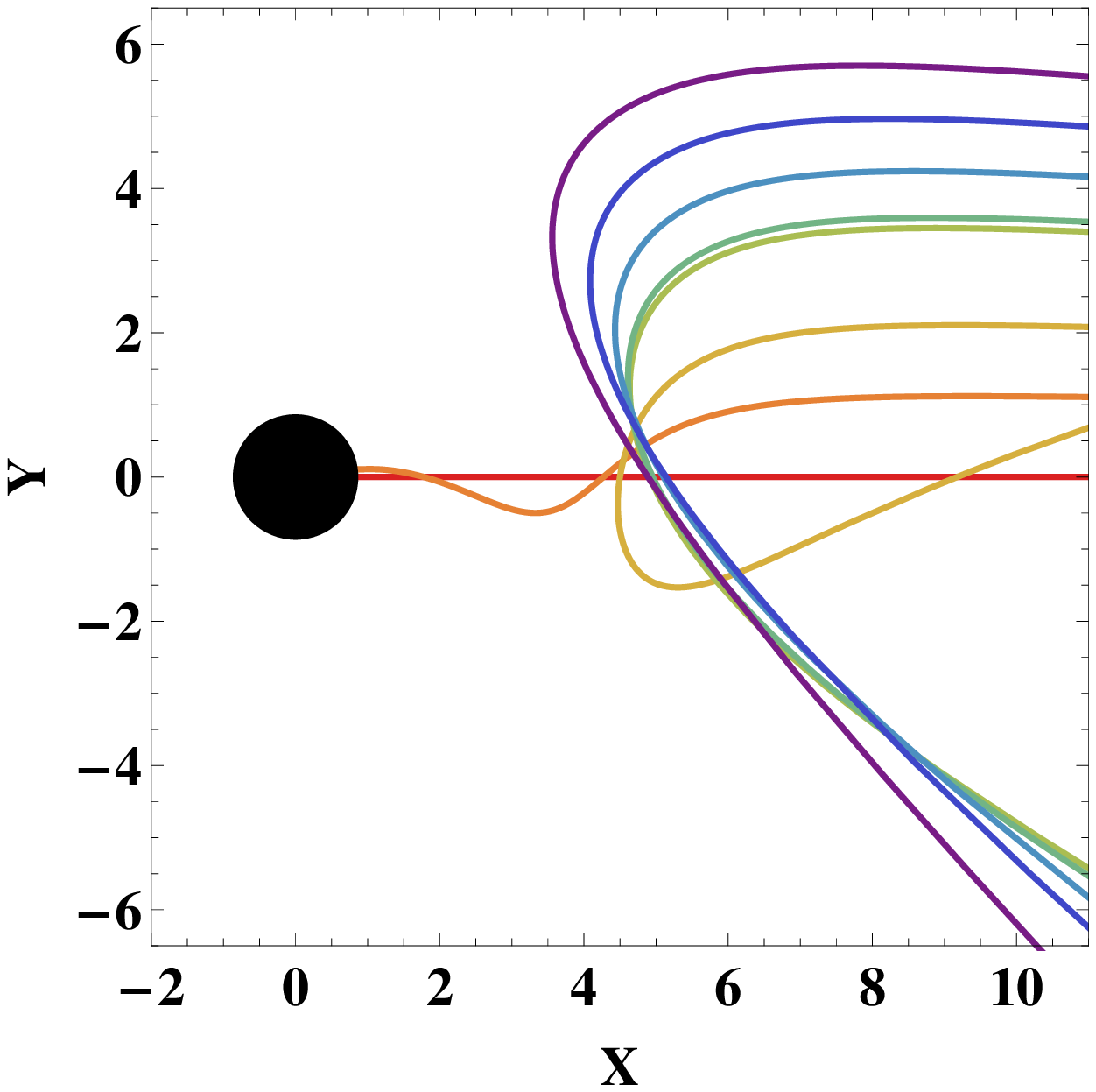}}
\subfigure[$\xi_{\psi}=0.997$]{\includegraphics[width=4cm]{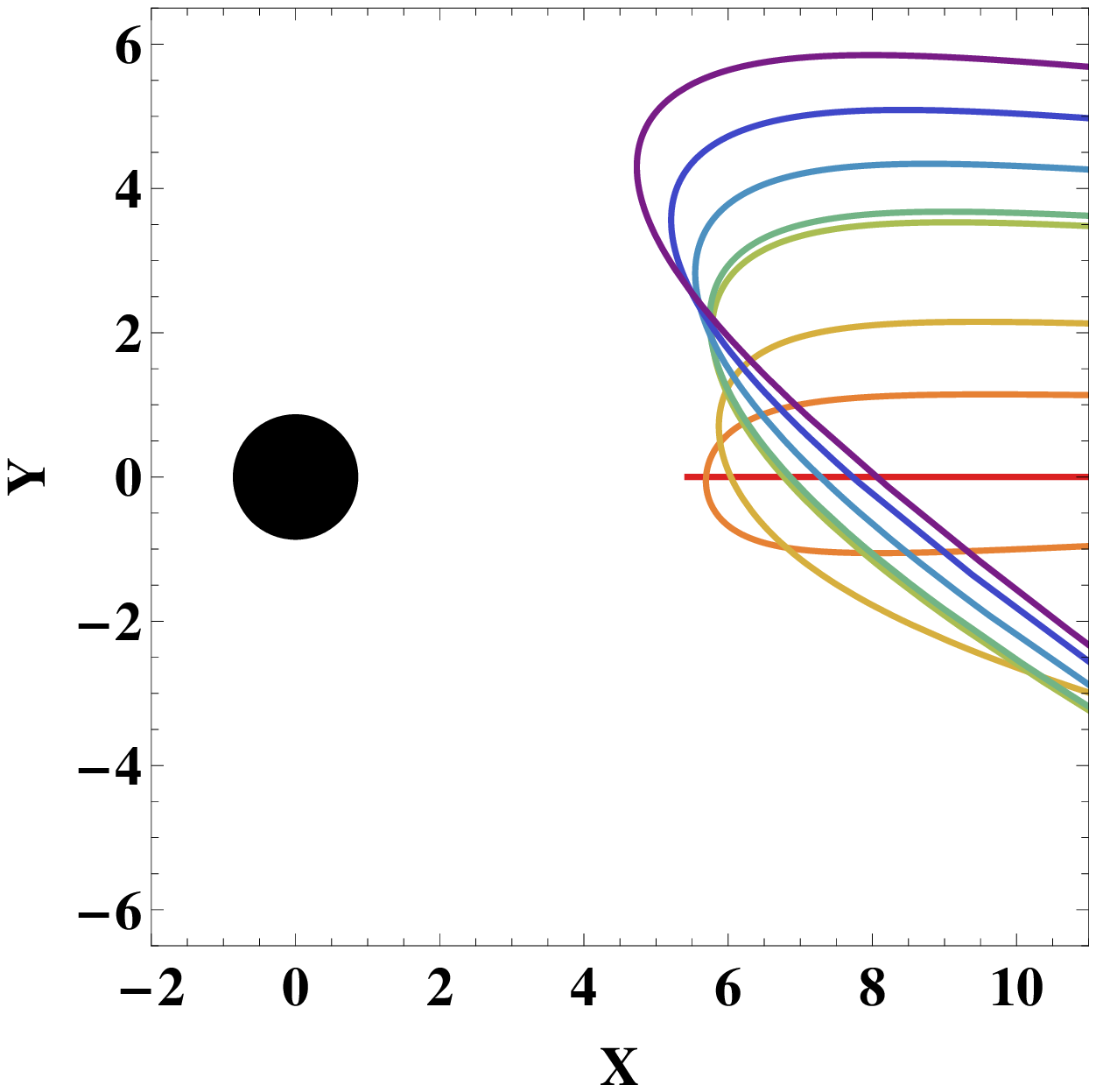}}
\caption{Propagation of photon with different $\xi_{\psi}$ in the rotating squashed KK black hole spacetime with the fixed $\rho_0=0.6$ and $b=0.4$. Here we set $\rho_M=1$, $X=\rho \sin\theta\cos\phi$ and $Y=\rho (\sin\theta\cos\phi+\cos\theta)/\sqrt{2}$. }\label{fig1s2}
\end{center}
\end{figure}
The mathematical reason is that the factor in the right side of Eq. (\ref{coordinate10}) $D(\rho_{ps})
-2H(\rho_{ps})\xi_{\psi}-A(\rho_{ps})\xi^2_{\psi}=0$ in this case. Here, we also present the propagation of photon with different $\xi_{\psi}$ in the rotating squashed KK black hole spacetime with the parameters ($\rho_0=0.6$, $b=0.2$)  in Fig. \ref{fig1s1} and ($\rho_0=0.6$,$b=0.4$) in Fig. \ref{fig1s2}, respectively. In Fig. \ref{fig1s1}, we find that the rotating squashed KK black hole with parameters $\rho_0=0.6$ and $b=0.2$ can capture the photons with various values of $\xi_{\psi}$  as they approach the black hole, which means that   black hole shadow exists in this case. However, for the black hole with parameters $\rho_0=0.6$ and $b=0.4$, we find from Fig. \ref{fig1s2} that for $\xi_{\psi}\geq 0.9 $, the impact parameter of the photon captured by black hole first increases and then decreases with increase of $\xi_{\psi}$, which is consistent with the change of $R_s$ in Fig. \ref{fig1}(c). Especially,
as $\xi_{\psi}\in(\xi_{\psi_c}, \frac{r'_{\infty}}{2})$, where $\xi_{\psi_c}$ is the positive root of the equation $D(\rho_{ps})-2H(\rho_{ps})\xi_{\psi}-A(\rho_{ps})\xi^2_{\psi}=0$, we find that the photons near black hole change their propagation direction and then become far away from the black hole. This implies that these photons can not captured by black hole so that they could reach the observer, which yields that there is no shadow in this case. Therefore, the specific angular momentum $\xi_{\psi}$ of photon from the fifth dimension plays an important role in the formation of no shadow for a rotating squashed KK black hole. The phenomenon of black hole without black shadow would vanish if there exists the further constraint on the specific angular momentum $\xi_{\psi}$ of photon from the fifth dimension.
\begin{figure}[ht]
\begin{center}
\includegraphics[width=6cm]{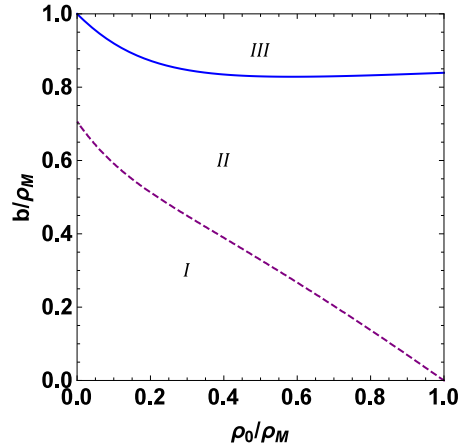}
\caption{ The parameter regions of existence and nonexistence of black hole shadow for a rotating squashed KK black hole. The black hole shadow exists only in the region $I$ and there is no shadow in the region $II$. In the  region  $III$, there is no horizon and the metric (\ref{metric1}) does not describe geometry of a black hole. Here  we set $\rho_M=1$. }\label{fig2}
\end{center}
\end{figure}
In Fig. \ref{fig2}, we present the parameter regions of existence and nonexistence of black hole shadow for a rotating squashed KK black hole. The black hole shadow exists only in the region $I$ and there is no black shadow in the region $II$. In the  region  $III$, there is no horizon and the metric (\ref{metric1}) does not describe geometry of a black hole.
\begin{figure}[ht]
\begin{center}
\includegraphics[width=6cm]{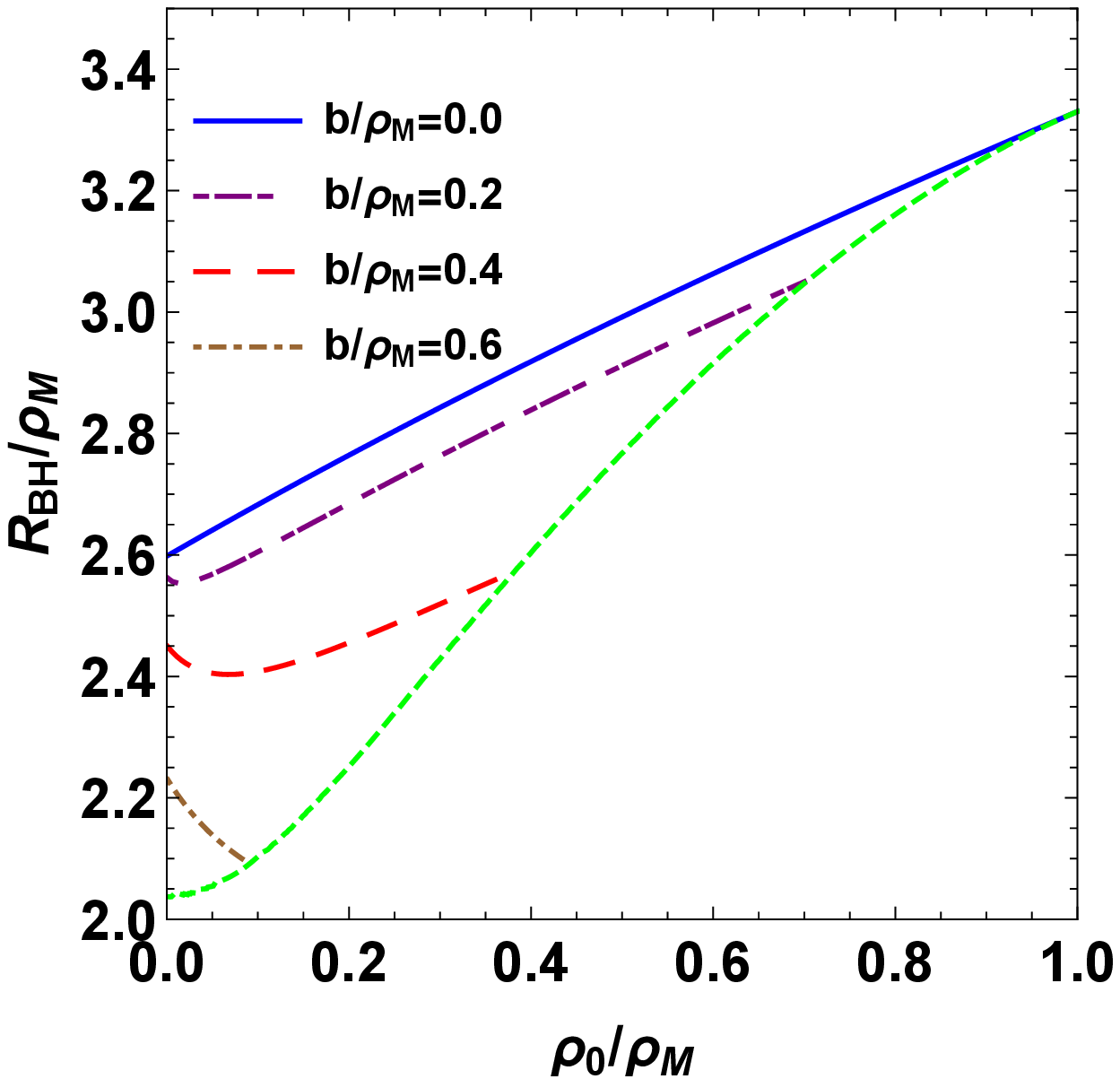}\quad\quad\includegraphics[width=6cm]{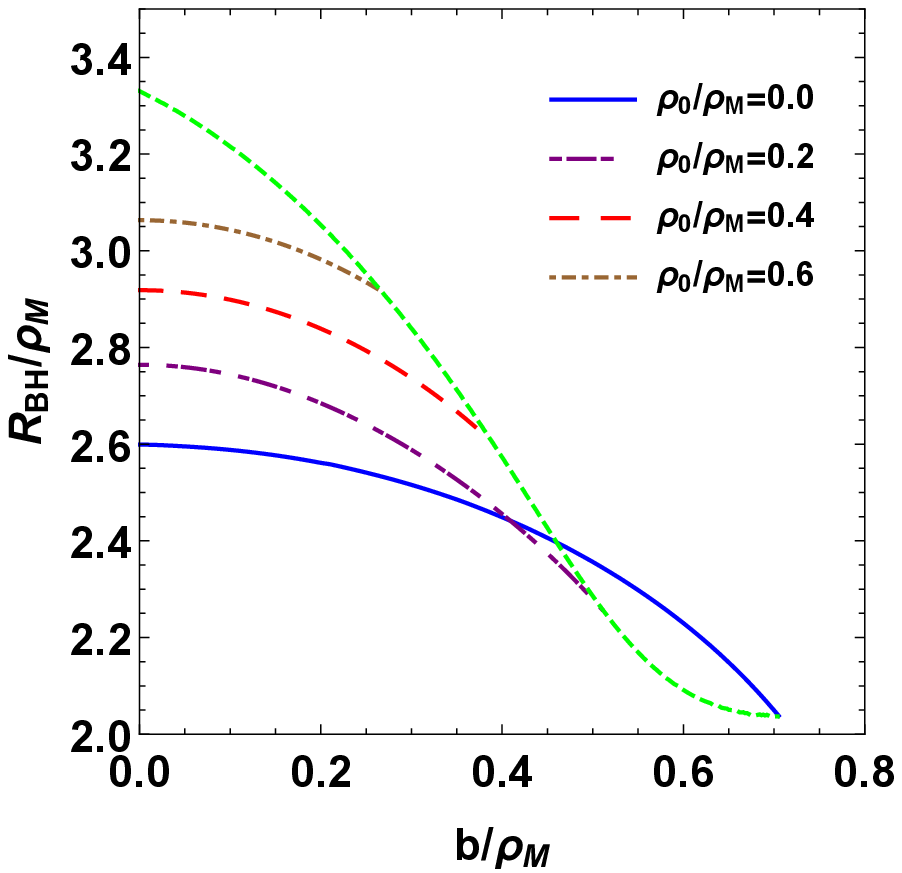}
\caption{ Dependence of the radius of the shadow on the rotation parameter $b$ and the scale of transition $\rho_0$ for a rotating squashed KK black hole. The green dashed line denote the boundary between existence and nonexistence of black hole shadow.}\label{fig3}
\end{center}
\end{figure}

In Fig. \ref{fig3}, we present the dependence of the radius of the shadow on the rotation parameter $b$ and the scale of transition $\rho_0$ in the case of  the spacetime parameters lie in the region $I$ in which there exists black hole shadow for a rotating squashed KK black hole. From Fig.\ref{fig3}, for the case with existence of black hole shadow, one can find that the  shadow radius $R_{BH}$ decreases with the rotation parameter $b$ of black hole, which is similar to those in the usual rotating black holes. With the increase of extra dimension parameter $\rho_0$, the radius $R_{BH}$ increases monotonically in the case of $b=0$.  In the rotating cases with the smaller $b$, we find that $R_{BH}$ first decreases and then increases with the parameter $\rho_0$. For the cases with the larger $b$, $R_{BH}$  decreases monotonically with the parameter $\rho_0$.

Finally, we make use of the metric (\ref{metric1}) and estimate the angular radius of the black hole shadow by using the observable $R_{BH}$ as $\theta_{BH}=R_{BH}\mathcal{M}/D_O$, where $D_O$ is the distance between the observer and the black hole. For an arbitrary black hole of mass $\mathcal{M}$ and distance $D_O$  from the observer, the angular radius can be expressed as $\theta_{BH}=9.87098\times10^{-6} R_{BH}(\mathcal{M}/M_{\odot})(1kpc/DO)\mu as$ \cite{w11,w16}
with  the mass of the Sun $M_{\odot}$. In Table (I), we present the angular radius of the black hole by the metric (\ref{metric1}) for the supermassive black hole
Sgr $A^{*}$ located at the Galactic center and the supermassive black hole in $M87$, respectively. Here we use the mass $\mathcal{M} = 4.3\times10^{6} M_{\odot}$ and the observer distance $D_O = 8.3 kpc$ for the black hole Sgr $A^{*}$\cite{Gillessen}, and $\mathcal{M} = 6.5\times10^{9} M_{\odot}$ and $D_O=16.8 Mpc$ for the black hole in the $M87$ \cite{e3}.
\begin{table}[ht]
\centering
\begin{tabular}{|c|c|c|c|c|c|c|c|c|c|}
\hline
\hline
\multicolumn{2}{|c|}{}&\multicolumn{4}{c|}{Sgr $A^{*}$}&\multicolumn{4}{c|}{ $M87$}\\
\cline{3-10}
\multicolumn{2}{|c|}{\multirow{2}{*}{$\theta_{BH}(\mu \ arcsec)$ }}&\multicolumn{4}{c|}{$b$ }&\multicolumn{4}{c|}{$b$ }\\
\cline{3-10}\multicolumn{2}{|c|}{}&$0.0$ &$0.2$&$0.4$&$0.6$&$0.0$ &$0.2$&$0.4$&$0.6$\\
\hline
\multirow{5}{*}{$\rho_{0}$ }& 0.0 &26.573  & 26.213 &  25.065 &22.829  &19.845  & 19.577 &  18.719 &17.050\\
\cline{2-10} & 0.2 &28.272 & 27.458 & 25.115  & $\cdot\cdot\cdot$   &21.114 & 20.506 & 18.756  & $\cdot\cdot\cdot$    \\
\cline{2-10} & 0.4 &29.850 & 29.035 &$\cdot\cdot\cdot$      &$\cdot\cdot\cdot$    &22.292 & 21.684 & $\cdot\cdot\cdot$       &$\cdot\cdot\cdot$       \\
\cline{2-10} & 0.6 &31.330 & 30.501 & $\cdot\cdot\cdot$      &$\cdot\cdot\cdot$   &23.397 & 22.778 & $\cdot\cdot\cdot$      &$\cdot\cdot\cdot$        \\
\cline{2-10} & 0.8 &32.729 & $\cdot\cdot\cdot$       & $\cdot\cdot\cdot$        &$\cdot\cdot\cdot$     &24.442 & $\cdot\cdot\cdot$       & $\cdot\cdot\cdot$        &$\cdot\cdot\cdot$    \\
\cline{2-10} & 1.0 &34.061 & $\cdot\cdot\cdot$      &$\cdot\cdot\cdot$      &$\cdot\cdot\cdot$     &25.437 & $\cdot\cdot\cdot$       & $\cdot\cdot\cdot$        &$\cdot\cdot\cdot$
\\ \hline\hline
\end{tabular}
\caption{The numerical estimation for the angular radius of the black shadow for the supermassive black hole Sgr A* in our Galaxy and the black hole in $M87$ by using the metric of a rotating squashed KK black hole.}
\label{table0}
\end{table}
The latest observation indicates that the angular diameter of $M87$ black hole is $42\pm3\mu as$ \cite{e3}. Combining it with the data in Table (\ref{table0}), one can find that there is a room for
the theoretical model of such a rotating squashed KK black hole.

\section{Summary and Discussion}

We have studied the shadow of a rotating squashed Kaluza-Klein black hole and find that the shadow possesses some novel properties differed from those of other rotating black holes. Firstly, the shadow shape is a perfect black disk for the black hole in the allowed parameter regions of $\rho_0$ and  rotation parameter $b$. It is different from that of the usual Kerr rotating black hole where the shape gradually changes from disk to ``D"-shape with the increase of rotation parameter. The circular silhouette of the shadow for the rotating squashed KK black hole is caused by the spacetime property that there are two equal rotational parameters in this special black hole spacetime, which also leads to that
the $\theta$- component  equation is independent of the rotation parameter. Moreover, since the radius $R_s$ of the image in the observer's sky caused by the photon falling into black hole horizon depends  on the specific angular momentum $\xi_{\psi}$ of photon, only the minimum value of $R_s$ is the radius of the black hole shadow, which yields that the shadow for a rotating squashed KK black hole is heavily influenced by the specific angular momentum $\xi_{\psi}$ of photon. Especially, as the black hole parameters lie in a certain special range, we find that there is no shadow for a black hole since the minimum value $R_{s_{\text{min}}}=0$ in these special cases, which is novel since it does not emerge in the usual black hole spacetimes. It must be noted that the black hole without shadow is not caused by that light rays can penetrate the black hole, but by that the photons near black hole with some special range of $\xi_{\psi}$ change their propagation direction and then become far away from the black hole.
Therefore, the specific angular momentum $\xi_{\psi}$ of photon from the fifth dimension plays an important role in the formation of no black shadow for a rotating squashed KK black hole. The phenomenon of black hole without black shadow would vanish if there exists the further constraint on the specific angular momentum $\xi_{\psi}$ of photon from the fifth dimension.
In the case where black hole shadow exists, we find that the radius of the black hole shadow $R_{BH}$ decreases with the rotation parameter $b$ of black hole, which is similar to those in the usual rotating black holes. With the increase of extra dimension parameter $\rho_0$, the radius of the black hole shadow $R_{BH}$ increases monotonically in the case with $b=0$. In the rotating cases with the smaller $b$, we find that $R_{BH}$ first decreases and then increases with the parameter $\rho_0$. For the cases with the larger $b$, $R_{BH}$  decreases monotonically with the parameter $\rho_0$.
Finally, we make use of the metric (\ref{metric1}) and estimate the angular radius of the black hole shadow by the observation data from the supermassive black hole
Sgr $A^{*}$ located at the Galactic center and the supermassive black hole in $M87$, which implies that there is a room for the theoretical model of such a rotating squashed KK black hole.

\section{Acknowledgments}

This work was partially supported by the National Natural Science Foundation of China under
Grant No. 11875026, the Scientific Research
Fund of Hunan Provincial Education Department Grant
No. 17A124. J. Jing's work was partially supported by
the National Natural Science Foundation of China under
Grant No. 11875025.

\end{document}